\documentclass[prb,reprint,superscriptaddress]{revtex4-1}

\usepackage{amsmath}
\usepackage{amssymb}

\usepackage{graphicx}
\graphicspath{ {figures/}, {figures_data/} }

\usepackage{hyperref}
\hypersetup{
 bookmarks=true,		% show bookmarks bar?
 unicode=false,			% nonLatin characters in Acrobat’s bookmarks
 pdftoolbar=true,		% show Acrobat’s toolbar?
 pdfmenubar=true,		% show Acrobat’s menu?
 pdffitwindow=false,		% window fit to page when opened
 pdfstartview={FitH},		% fits the width of the page to the window
 pdftitle={The correlation theory of the chemical bond},    % title
 pdfauthor={Szil{\'a}rd Szalay} {Gergely Barcza} {Tibor Szilv\'asi} {Libor Veis} {\"Ors Legeza},	% author
 pdfsubject={chemical bonds are identified with strong orbital correlation},	% subject of the document
 pdfcreator={pdflatex},		% creator of the document
 pdfproducer={vim},		% producer of the document
 pdfkeywords={quantum chemistry} {chemical bond} {covalent} {aromaticity} {quantum correlations} {entanglement} {multiorbital correlations}, % list of keywords
 pdfnewwindow=true,		% links in new window
 colorlinks=true,		% false: boxed links; true: colored links
 linktoc=page,			% defines which part of an entry in the table of contents is made into a link
 linkcolor=blue,		% color of internal links	(red)
 citecolor=blue,		% color of links to bibliography
 filecolor=blue,		% color of file links
 urlcolor=blue			% color of external links
}
% ...

\newcommand{\field}[1]{\mathbb{#1}}
\newcommand{\vs}[1]{\boldsymbol{#1}}
\newcommand{\cket}[1]{\vert #1 \rangle}
\newcommand{\bra}[1]{\langle #1 \vert}
\DeclareMathOperator{\tr}{tr}
\DeclareMathOperator*{\argmin}{argmin}

\DeclareMathOperator{\downset}{\downarrow}
\DeclareMathOperator{\upset}{\uparrow}

\def\precdot{\mathrel{%
   \mathchoice{\PRECDOT}{\PRECDOT}{\scriptsize\PRECDOT}{\tiny\PRECDOT}%		this is wrong in maht mode, but works, 
}}
\def\PRECDOT{{%
    \setbox0\hbox{$\prec$}%
    \rlap{\hbox to \wd0{\hss$\cdot$}}\box0
}}

\def\SUCCDOT{{%
    \setbox0\hbox{$\succ$}%
    \rlap{\hbox to \wd0{\hss$\cdot$}}\box0
}}

\newcommand{\finereq}{\preceq}
\newcommand{\nfinereq}{\npreceq}
\newcommand{\finer}{\prec}
\newcommand{\nfiner}{\nprec}
\newcommand{\covered}{\precdot}

\newcommand{\coarser}{\succ}

\providecommand{\abs}[1]{\lvert#1\rvert}
\providecommand{\bigabs}[1]{{\bigl\lvert#1\bigr\rvert}}

\newcommand{\set}[1]{\{ #1 \}}
\newcommand{\bigset}[1]{\bigl\{ #1 \bigr\}}

\newcommand{\sset}[2]{\{ #1 \;\vert\; #2 \}}
\newcommand{\bigsset}[2]{\bigl\{ #1 \;\big\vert\; #2 \bigr\}}

\begin{document}

\title{The correlation theory of the chemical bond}
% **********************
\author{Szil\'ard Szalay}
\email{szalay.szilard@wigner.mta.hu}
\affiliation{Strongly Correlated Systems ``Lend{\"u}let'' Research Group,
Institute for Solid State Physics and Optics,
MTA Wigner Research Centre for Physics,
H-1121 Budapest, Konkoly-Thege Mikl{\'o}s {\'u}t 29-33, Hungary}
% **********************
\author{Gergely Barcza}
\email{barcza.gergely@wigner.mta.hu}
\affiliation{Strongly Correlated Systems ``Lend{\"u}let'' Research Group,
Institute for Solid State Physics and Optics,
MTA Wigner Research Centre for Physics,
H-1121 Budapest, Konkoly-Thege Mikl{\'o}s {\'u}t 29-33, Hungary}
% **********************
\author{Tibor Szilv\'asi}
\email{szilvasitibor@ch.bme.hu}
%\email{tib.szilvasi@gmail.com}
\affiliation{Department of Chemical and Biological Engineering,
University of Wisconsin-Madison,
1415 Engineering Drive, Madison, Wisconsin 53706, United States}
\affiliation{Department of Inorganic and Analytical Chemistry,
Budapest University of Technology and Economics,
H-1111 Budapest, Szent Gell\'ert t\'er 4, Hungary}
% **********************
\author{Libor Veis}
\email{libor.veis@jh-inst.cas.cz}
\affiliation{J. Heyrovsk\'y Institute of Physical Chemistry,
Academy of Sciences of the Czech Republic,
CZ-18223 Prague, Czech Republic}
% **********************
\author{\"Ors Legeza}
\email{legeza.ors@wigner.mta.hu}
\affiliation{Strongly Correlated Systems ``Lend{\"u}let'' Research Group,
Institute for Solid State Physics and Optics,
MTA Wigner Research Centre for Physics,
H-1121 Budapest, Konkoly-Thege Mikl{\'o}s {\'u}t 29-33, Hungary}

%%%%%%%%%%%%%%%%%%%%%%%%%%%%%%%%%%%%%%%%%%%%%%%%%%%%%%%%%%%%%%%%%%%%%%%%%%%%%%%%
\begin{abstract}
The quantum mechanical description of the chemical bond
is generally given in terms of
delocalized bonding orbitals,
or, alternatively,
in terms of correlations of occupations of localised orbitals.
However, in the latter case, multiorbital correlations were treated only in terms of two-orbital correlations,
although the structure of multiorbital correlations is far richer;
and, in the case of bonds established by more than two electrons, 
multiorbital correlations represent a more natural point of view.
Here, for the first time,
we introduce the true multiorbital correlation theory,
consisting of
a framework for handling the structure of multiorbital correlations,
a toolbox of true multiorbital correlation measures,
and the formulation
of the multiorbital correlation clustering,
 together with an algorithm for obtaining that.
These make it possible to characterise quantitatively,
how well a bonding picture describes the chemical system.
As proof of concept,
we apply the theory for the investigation of the bond structures of several molecules.
We show that 
the non-existence of well-defined multiorbital correlation clustering
provides a reason for debated bonding picture. 

\end{abstract}

\maketitle

%%%%%%%%%%%%%%%%%%%%%%%%%%%%%%%%%%%%%%%%%%%%%%%%%%%%%%%%%%%%%%%%%%%%%%%%%%%%%%%%
%%%%%%%%%%%%%%%%%%%%%%%%%%%%%%%%%%%%%%%%%%%%%%%%%%%%%%%%%%%%%%%%%%%%%%%%%%%%%%%%
\section*{Introduction}

Since quantum theory is a probabilistic theory,
it is not surprising that
using concepts of quantum information theory\cite{Wilde-2013,Nielsen-2000} 
turns out to be fruitful in several fields of research in which quantum theory is involved.
Maybe the most important notion in a probabilistic theory is
correlation,\cite{Amico-2008}
and, in quantum systems, also entanglement.\cite{Horodecki-2009,Modi-2010}
Taking their investigation as a guiding principle
has already led to
important achievements
in several fields of research,\cite{Legeza-2003b,Legeza-2006a,Rissler-2006,Amico-2008}
recently also in quantum chemistry.\cite{Legeza-2003b,Legeza-2006a,Rissler-2006,Pipek-2009,Barcza-2011,McKemmish-2011,Boguslawski-2013,Kurashige-2013,Mottet-2014,Fertitta-2014,Duperrouzel-2014,Boguslawski-2015,Freitag-2015,Szilvasi-2015,Zhao-2015}

The notion of chemical bond\cite{Lewis-1916} is a very useful concept in chemistry.
It originated at the beginning of chemistry,
it is expressive for the classically thinking mind,
and the errors arising from the approximative nature of the concept can often be ignored.
In the first half of the twentieth century, however,
we learned that the proper description of the microworld is given by quantum mechanics.
Quantum mechanics gives more accurate results for chemical systems
than any preceding model,
however, it is very inexpressive for the classically thinking mind.
One of the most used quantum mechanical concepts of the chemical bond is the valence bond theory,\cite{Shaik-2007}
 among others,\cite{Bader-1975,Daudel-1975} 
forming the bonds between atoms by overlap of the atomic orbitals. 
The valence bond theory complements the molecular orbital theory,\cite{Fleming-2010}
distributing pairs of electrons in bonding molecular orbitals delocalized over the system. 
In this work, in the spirit of the valence bond theory, we study correlations among the orbitals localised on individual atoms.

Indeed, studying the \emph{two-orbital} correlation pattern
in molecular systems
in equilibrium
gives us the hint that
the correlations must be related to the chemical bonds:
strong two-orbital correlations can be observed between the orbitals 
which are involved in the given bond.%
\cite{Rissler-2006,Barcza-2011,Boguslawski-2013,Kurashige-2013,Mottet-2014,Fertitta-2014,Duperrouzel-2014,Boguslawski-2015,Freitag-2015,Szilvasi-2015,Murg-2015,Zhao-2015}
Simple covalent bonds formed by two atomic orbitals fit well into this \emph{two-orbital correlation} picture.
However, there are more complicated bonding scenarios with electrons shared by multiple atoms,
in this case some true \emph{multiorbital correlation} picture should be used.\cite{Szalay-2015b}
(So far, multiorbital correlations were investigated by the use of the notion of two-orbital correlation only.%
\cite{Legeza-2006a,Rissler-2006,Barcza-2011,Boguslawski-2013,Kurashige-2013,Mottet-2014,Fertitta-2014,Duperrouzel-2014,Boguslawski-2015,Freitag-2015,Szilvasi-2015,Murg-2015,Barcza-2015,Zhao-2015}%
)
The reason for this is twofold.
On the one hand, such bonds, e.g., a delocalized ring in a benzene molecule,
cannot be considered as a ``sum'' of two-orbital bonds, but a true multiorbital bond.
On the other hand, 
in multiorbital systems,
\emph{hidden correlations} may occur,
that is,
there may be strong multiorbital correlation among the orbitals in a cluster
even if the two-orbital correlations are weak.

In this work we provide the true multiorbital correlation theory,
consisting of
a framework for handling the structure of multiorbital correlations,
a toolbox of true multiorbital correlation measures,
and the definition together with an algorithm for the multiorbital correlation clustering.
The presented theory significantly outgrows the multipartite entanglement theory,\cite{Szalay-2015b}
on which it is based, namely, in the last three items mentioned just above.
(The detailed construction is presented in the Appendix.)
We adopt the principle that
bonds are where the electrons can freely move among atoms,
and this is reflected in the correlations of occupations of localised orbitals.
Then we show illustrative results by
investigating the multipartite correlations
in several molecules showing multiorbital bonds.
We give quantitative characterisation how well a bonding picture describes the chemical systems.
We also illustrate that in the debated case of the dicarbon molecule,
there is no well-defined multiorbital correlation clustering,
which provides a reason for the ambiguous bonding picture.\cite{Shaik-2012,Shaik-2013,Grunenberg-2012,Frenking-2012,Zhong-2016}
This is not only the first true multiorbital correlation based study of the chemical bond,
but also the first application of true multipartite correlation based techniques in physics.

We emphasise that the notions of correlations are basis dependent.
We employed two basis sets in this study, standard STO-3G and STO-6G with optimised exponents,
which latter provided HF energy close to HF/cc-pVTZ level of theory (see Methods section),
together with the localisation procedure of Pipek-Mezey\cite{Pipek-1989} to produce atomic-like orbitals.
All results discussed about the correlation structure of localised orbitals 
are understood with respect to this localisation.
For the prototypical molecules which were considered for illustrating our theory,
we have found that employing the minimal unchanged STO-3G basis set is sufficient for the description of bonding,
and using a basis set closer to the Complete Basis Set limit have not changed the bonding picture.
The results using optimised STO-6G basis set are presented in the main text,
while the results using unchanged STO-3G basis set are also presented in the Appendix for comparison.

We note that our work is not connected to previous works of de Giambiagi, Giambiagi and Jorge\cite{deGiambiagi-1985}
regarding generalised bond indices based on density-density correlation functions.

%%%%%%%%%%%%%%%%%%%%%%%%%%%%%%%%%%%%%%%%%%%%%%%%%%%%%%%%%%%%%%%%%%%%%%%%%%%%%%%%
%%%%%%%%%%%%%%%%%%%%%%%%%%%%%%%%%%%%%%%%%%%%%%%%%%%%%%%%%%%%%%%%%%%%%%%%%%%%%%%%
\section*{Multiorbital correlations}

For the quantum mechanical description of the molecule,
we use the second quantized picture,
that is, the Hilbert space of the electronic system 
is built up by the one-orbital Hilbert spaces,
describing the occupation of the orbitals.\cite{Szalay-2015a}
An orbital can be unoccupied, occupied with one electron of spin up or down, or doubly occupied by two electrons of spin up and down,
resulting in four-dimensional one-orbital Hilbert spaces.
In the Hilbert space formalism of quantum mechanics,
any linear combination of orbitals is an orbital,
however, the interpretation, or physical properties single out some of them.
Correlations among orbitals are not invariant under such nonlocal (among-orbital) operations.
In order that the correlations express some connection among local objects (atoms),
it is necessary that the orbitals are \emph{localised} on the atoms.

So, for the description of (the electronic system of) the \emph{molecule,}
we consider $m$ localised, atomic-like orbitals.
Let $M$, stands for ``molecule'', denote the set of (the labels of) these orbitals. 
We aim at describing the correlations in an $L\subseteq M$ set of orbitals (cluster).
(If $L=M$ then the correlations in the whole molecule is considered.)
In general, the \emph{state} of the full electronic system of the cluster $L$ can be described by the density operator\cite{Ohya-1993,Araki-2003b,Wilde-2013} $\varrho_L$.
The \emph{reduced state} of a (sub)cluster $X\subseteq L$ can be described by the reduced density operator\cite{Ohya-1993,Araki-2003b,Wilde-2013}
$\varrho_X$.
If the cluster of orbitals $L$ can be described by a state vector $\cket{\psi_L}$
(for example, when a given eigenstate of the whole molecule is considered),
then its density operator is of rank one, $\varrho_L=\cket{\psi_L}\bra{\psi_L}$, called a \emph{pure state}.
Its reduced density operator is mixed (not of rank one) in general,
which is the manifestation of the \emph{entanglement}\cite{Horodecki-2009} of (sub)cluster $X$ and the rest of the cluster $L\setminus X$.

The correlation can be defined with respect to a split of the $M$ set of the orbitals.\cite{Szalay-2012,Szalay-2015b}
Let $\xi$ denote such a split, that is, a \emph{partition}\cite{Davey-2002}
$\xi =\{X_1,X_2,\dots,X_{\abs{\xi}}\} \equiv X_1|X_2|\dots|X_{\abs{\xi}}$,
where the clusters $X\in\xi$, called \emph{parts}, are disjoint subsets of the cluster $L$,
and their union is the full cluster $L$.
A natural comparison of partitions is the ``refinement'':
we say that partition $\upsilon$ is finer than partition $\xi$,
if the parts of $\upsilon$ are contained in the parts of $\xi$.
The set of the partitions of $L$ is denoted with $\Pi(L)$.
(For illustrations, see Fig.~\ref{fig:partlattices} online.)
The measure of correlation with respect to this split is the
\emph{$\xi$-correlation},\cite{Szalay-2015b}
\begin{equation}
\label{eqr:xiCorr}
C_\xi(\varrho_L) := \sum_{X\in\xi} S(\varrho_X) - S(\varrho_L).
\end{equation}
Here $S(\varrho)=-\tr\varrho\ln\varrho$ is the von Neumann entropy.\cite{Ohya-1993,Wilde-2013}
As a special case, the $i|j$-correlation
\begin{equation}
\label{eqr:twoorbitMutInf}
C_{i|j}(\varrho_{\{i,j\}}) = S(\varrho_{\{i\}}) + S(\varrho_{\{j\}}) - S(\varrho_{\{i,j\}}) = I_{i|j}(\varrho_{\{i,j\}}),
\end{equation}
being the well-known (two-orbital) mutual information,\cite{Ohya-1993,Wilde-2013,Adesso-2016}
has already been used.%
\cite{Legeza-2006a,Rissler-2006,Barcza-2011,Boguslawski-2013,Kurashige-2013,Mottet-2014,Fertitta-2014,Duperrouzel-2014,Boguslawski-2015,Freitag-2015,Szilvasi-2015,Murg-2015,Barcza-2015,Zhao-2015}
The $\xi$-correlation is zero if 
the state is uncorrelated with respect to $\xi$ 
(it can be written in a product form of reduced states of clusters $X\in\xi$);
and nonzero otherwise,
characterising the strength of the correlation among the parts $X\in\xi$.
This comes from the information-geometrical meaning of this quantity:
it characterises how ``far'' the state is from the states uncorrelated with respect to $\xi$.
(For more details of the construction, see the Appendix.)
Note that $C_\xi$ is larger for finer partitions,
(this is called \emph{multipartite monotonicity}\cite{Szalay-2015b}),
it is zero for the trivial split $\xi=\top=\{L\}$,
and it takes its maximum, $C_\bot$, for the finest split $\xi = \bot = \{ \{i\} \;\vert\; i\in L \}$.
The latter quantity
is also called \emph{total correlation},\cite{Lindblad-1973,Horodecki-1994,Legeza-2004b,Legeza-2006b,Herbut-2004} 
\begin{equation}
\label{eqr:totalCorr}
C_\text{tot}(\varrho_L) := C_\bot (\varrho_L)=\sum_{i\in L}S(\varrho_{\{i\}})-S(\varrho_L).
\end{equation}
(Note that if cluster $L$ is described by a pure state, e.g., if $L=M$, then $S(\varrho_L)=0$,
and the correlation is entirely quantum entanglement.\cite{Horodecki-2009,Modi-2010,Szalay-2015b})
It is easy to check the following \emph{sum rule},\cite{Herbut-2004} valid for any partition $\xi$,
\begin{equation}
\label{eqr:totalCorrpart}
\sum_{X\in\xi} C_{\bot,X}(\varrho_X) + C_\xi(\varrho_L) = C_\bot(\varrho_L),
\end{equation}
that is, the total correlation
is the sum of the total correlations inside the parts plus the
correlation with respect to the partition.

We would also like to characterise the correlations
in an overall sense, that is, without respect to a given partition.
There are several ways of this,\cite{Szalay-2015b}
here we consider two of them.
Let us introduce 
the \emph{$k$-partitionability correlation} and
the \emph{$k$-producibility correlation}, respectively,
\begin{subequations}
\label{eqr:kppCorr}
\begin{align}
C_\text{$k$-part}(\varrho_L) &:= \min_{\xi:\; \abs{\xi}\geq k}C_\xi(\varrho_L),\qquad\qquad\\
C_\text{$k$-prod}(\varrho_L) &:= \min_{\xi:\; \forall X\in\xi,\: \abs{X}\leq k }C_\xi(\varrho_L),
\end{align}
\end{subequations}
for $1\leq k\leq\abs{L}$.
These characterise two different (one-parameter-) notions of multiorbital correlations.
The $k$-partitionability correlation is zero if
the cluster can be split into at least $k$ parts which are uncorrelated with one another,
and the correlations are restricted only inside those parts;
and nonzero otherwise,
characterising the strength of this kind of correlation.
In general, $C_{\text{$k$-part}}$ increases with $k$,
and it jumps after the number $k$ of parts
into which $L$ can approximately be split.
The $k$-producibility correlation is zero if
the cluster $L$ contains correlated (sub)clusters of size not larger than $k$;
and nonzero otherwise,
characterising the strength of this kind of correlation.
In general, $C_{\text{$k$-prod}}$ decreases with $k$,
and it jumps at the size $k$ of the largest part
in the partition into which $L$ can approximately be split.
As special cases, 
$C_\text{$\abs{L}$-part}=C_\text{$1$-prod}=C_\bot$ grabs all the correlations,
it is zero if there is no correlation at all in the cluster $L$,
that is, its state is a product of the states of orbitals;
and nonzero otherwise.
On the other hand,
$C_\text{$2$-part}=C_\text{$(\abs{L}-1)$-prod}$ is sensitive only for the strongest correlations,
it is nonzero if the cluster $L$ is globally correlated,
that is, its state is not a product of states of two (or more) clusters;
and zero otherwise.
($C_\text{$1$-part}=C_\text{$\abs{L}$-prod}=C_\top = 0$.
For other values of $k$ there are no such coincidences among the partitionability and producibility correlations,
however, the relation $C_\text{$k$-part}\geq C_\text{$(\abs{L}-k+1)$-prod}$ holds.
Also, the bounds
$C_\text{$k$-part}\leq 2(k-1) (\ln 4)$,
$C_\text{$k$-prod}\leq 2(\abs{L}-k) (\ln 4)$ hold.
For more details, see the Appendix.)

The tools \eqref{eqr:xiCorr} and \eqref{eqr:kppCorr}, despite being so simple,
are proven very useful in a wide range of applications
for the characterisation
of multiorbital correlations in the electronic system of molecules.
In the sequel, we show four of these applications.
Illustrating these, we present numerical results for several prototypical molecules, 
namely benzene, pyrrole, borole, cyclobutadiene, furan, thiophene, and the sequence $\mathrm{C}_2\mathrm{H}_{2x}$ for $x=1,2,3$ and $\mathrm{C}_2$.

\begin{figure*}
\centering
\includegraphics{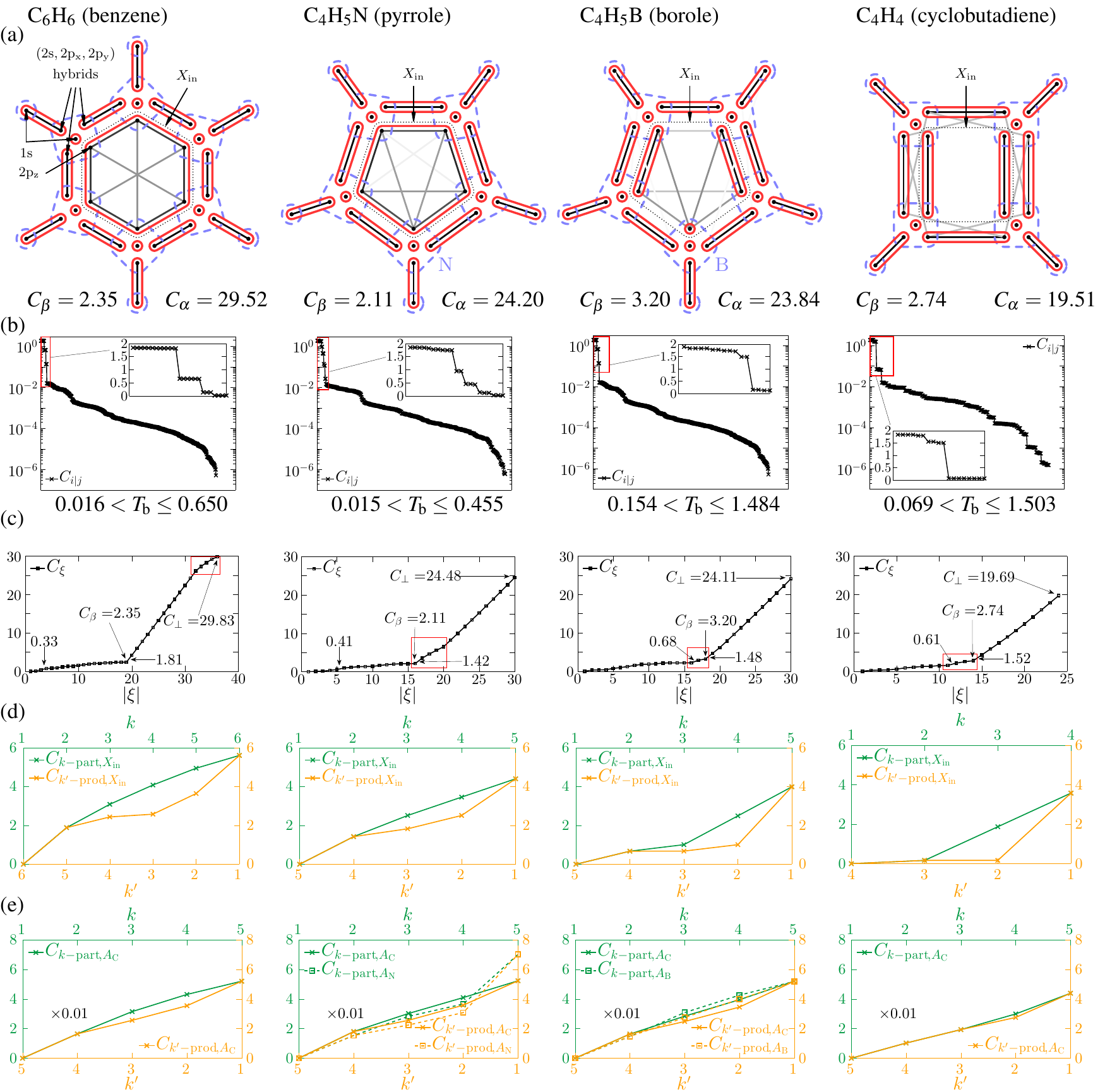}
\caption{Partitioning and multipartite correlations for the benzene, pyrrole, borole and cyclobutadiene molecules.
(a) Schematic view of the molecules:
the dots represent atomic orbitals,
the ones localised on an atom are encircled in dashed blue lines,
this is the atomic split $\alpha$,
the ones strongly correlated with each other are encircled in solid red lines,
this is the bond split $\beta$.
Strength of edges represent two-orbital correlations (shaded by a logarithmic scale).
The correlations $C_\alpha$ and $C_\beta$ are also shown.
(b) The distributions of the two-orbital correlations.
The possible ranges of two-orbital correlation thresholds $T_\text{b}$ are also shown.
(c) The $C_\xi$ tendencies of the successive bipartitioning.
The humps arising from the bipartitioning of multiorbital correlated clusters are indicated with red frames.
The maximal step before $\beta$ and the minimal step following $\beta$ are also shown.
(d) The correlations $C_{\text{$k$-part},X_\text{in}}$, $C_{\text{$k$-prod},X_\text{in}}$ for the inner bonding ($2\mathrm{p}_\mathrm{z}$) orbitals, contained in $X_\text{in}$.
(e) The correlations $C_{\text{$k$-prod},A}$, $C_{\text{$k$-part},A}$ for selected atoms A.
(The numerical values of the correlation measures are given in units of $\ln4$.)
}
\label{fig:resultscyclic}
\end{figure*}

\begin{figure}
\centering
\includegraphics{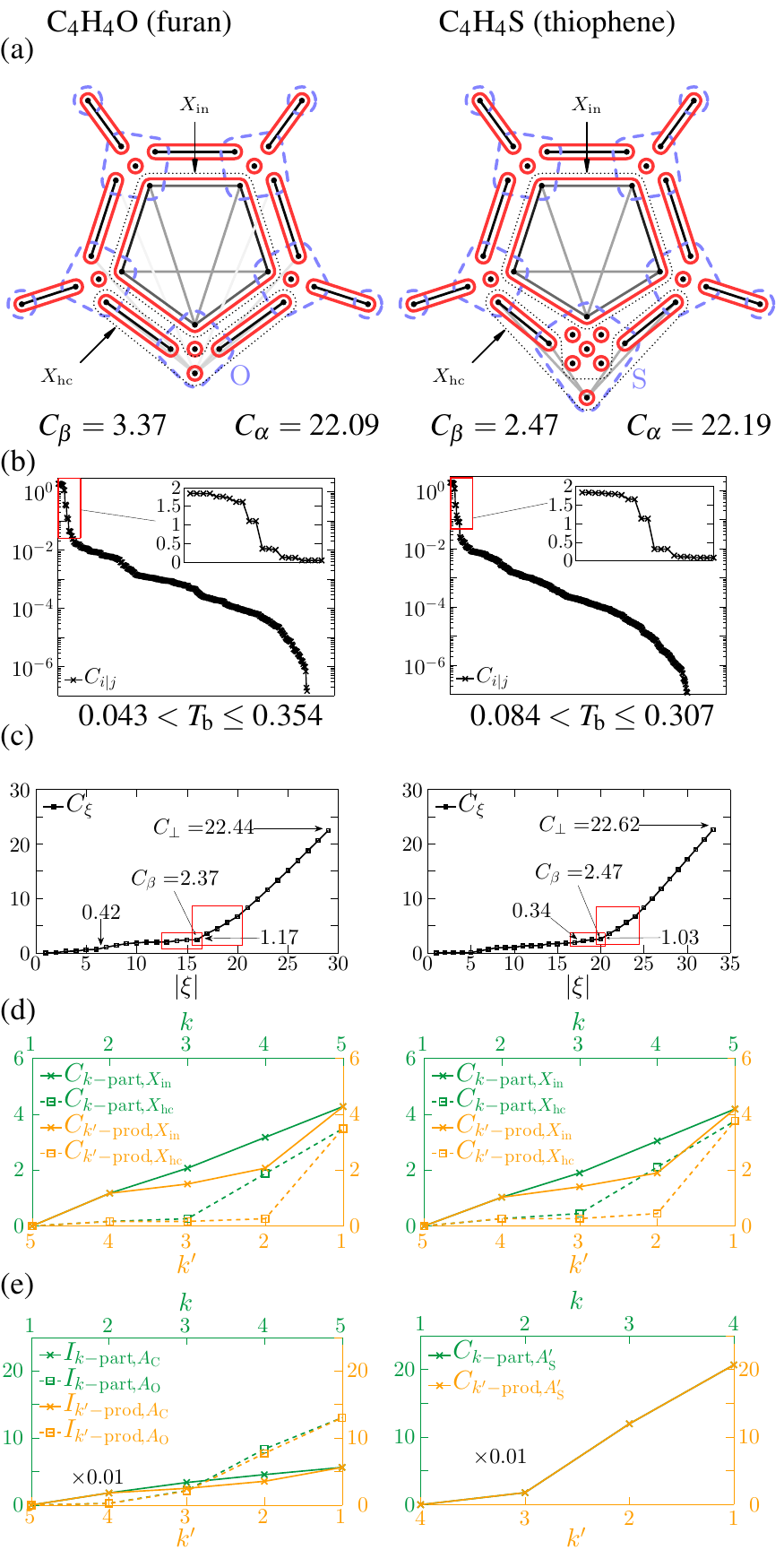}
\caption{Partitioning and multipartite correlations for the furan and thiophene molecules.
The same types of data are shown as in Fig.~\ref{fig:resultscyclic}.
(d) The correlations %$C_{\text{$k$-part},X_\text{hc}}$, $C_{\text{$k$-prod},X_\text{hc}}$ 
for the orbitals participated in the hyperconjugative interaction, contained in $X_\text{hc}$ are also shown.
(e) For the thiophene, the correlations among the valence orbitals, contained in $A_{\textrm{S}}'$ are shown.
}
\label{fig:resultscyclic2}
\end{figure}

\begin{figure*}
\centering
\includegraphics{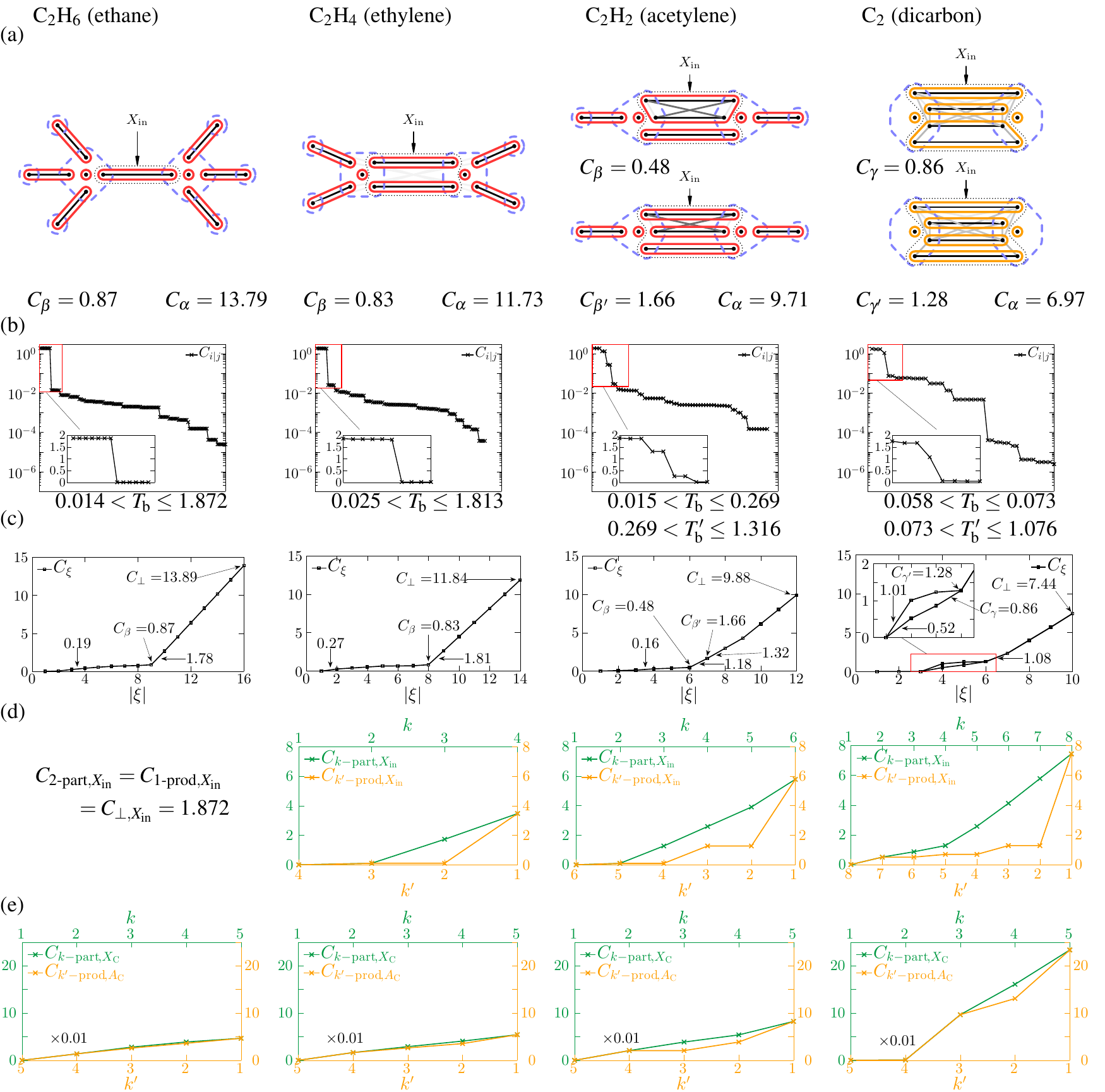}
\caption{Partitioning and multipartite correlations for the $\mathrm{C}_2\mathrm{H}_{2x}$ molecules.
The same types of data are shown as in Fig.~\ref{fig:resultscyclic}.}
\label{fig:resultsC2H2x}
\end{figure*}

%%%%%%%%%%%%%%%%%%%%%%%%%%%%%%%%%%%%%%%%%%%%%%%%%%%%%%%%%%%%%%%%%%%%%%%%%%%%%%%%
%%%%%%%%%%%%%%%%%%%%%%%%%%%%%%%%%%%%%%%%%%%%%%%%%%%%%%%%%%%%%%%%%%%%%%%%%%%%%%%%
\section*{Applications}
\subsection*{Application 1: Molecule, formed by bonds}

Our fundamental \emph{principle} is that,
in the equilibrium,
\textit{the bonds are almost uncorrelated with one another, 
and the orbitals involved in a bond are strongly (multiorbital) correlated.}
Using the tools introduced above,
we formulate this principle, 
and we demonstrate it for the aforementioned molecules.

An \emph{ansatz} for the bond structure
is given by a partition $\beta=B_1|B_2|\dots|B_{\abs{\beta}}\in\Pi(M)$ (\emph{bond split}),
representing the molecule as a set of \emph{bonds}
(represented by $B\subseteq M$ sets of orbitals),
together with some nonbonding orbitals (e.g., core orbitals or lone pairs, for those, $\abs{B}=1$).
Then the $\beta$-correlation $C_\beta(\varrho_M)$, given in \eqref{eqr:xiCorr},
characterising the correlation with respect to the ansatz $\beta$,
expresses how well this ansatz describes the physical situation:
the lower the $C_\beta$ the better the ansatz from a \emph{purely information-theoretical point of view.}
The aim of this application is to find the bond split $\beta$ (if exists) from \textit{ab initio} data,
without taking into account anything which can a priory be known about the bond structure in quantum chemistry.
We call this \emph{multiorbital correlation clustering.}

Since in a real electronic system
one cannot expect such a simple ansatz to be exactly valid ($C_\beta=0$), 
we actually pose the question,
which ansatz $\beta$ is the best choice for the description of the bonds 
from a \emph{physical chemical point of view.}
Being the best, however, is a delicate question.
Note that, on the one hand, 
$C_\top=0$, for the trivial split $\beta=\top$, which takes the whole molecule to be one big bond.
On the other hand, $C_\beta$ grows with respect to the refinement of $\beta$, 
and takes its maximal value $C_\bot$, the total correlation \eqref{eqr:totalCorr},
for the finest split $\beta = \bot$, which excludes nontrivial bonds.
These extremal cases, obviously, do not give proper descriptions
of the bond structure of a molecule,
since, on the one hand, there can be clusters weakly correlated but not uncorrelated with the remaining part of the molecule,
on the other hand, it is not allowed to neglect strong correlations.
Instead of these, we have to be able to split the molecule into 
\emph{weakly} correlated clusters consisting of \emph{strongly} correlated orbitals.
In order to grasp the meaning of the multipartite correlation clustering,
we have to be able to decide
about a given $\xi$,
if it is a good ansatz, or it is worth considering a $\xi'$,
which is ``a bit'' finer than $\xi$.
That is, 
we have to investigate the difference $C_{\xi'}-C_\xi$,
where $\xi'$ is finer than $\xi$, and there is no other partition between them.
We seek $\beta$,
for which,
\textit{while $\xi$ is coarser than $\beta$,
this difference is small if $\xi'$ is coarser than $\beta$,
but large, if $\xi'$ is not coarser than $\beta$.}
If there exists such a $\beta$, then it is meaningful to consider the electronic system
to be weakly correlated bonds consisting of strongly correlated orbitals,
and this is described by $\beta$.
(For the whole construction, see the Appendix.)

Here we face the problem that
verifying that this definition holds for a given partition
(calculating $S(\varrho_X)$ for all clusters $X\subseteq M$,
needed for the calculation of $C_\xi$ for all partitions $\xi$) 
is numerically prohibitive.
We can decrease the demands by
successive refinement of the partitioning (bipartitioning of one cluster in each step)
following the smallest increase in $C_\xi$.
One can show that 
if there exists a $\beta$ satisfying the definition above, 
then the successive refinement goes through $\beta$,
$C_\xi$ increases slowly until $\beta$, and rapidly after $\beta$.
(For the whole construction, see the Appendix.)

We can also have a hint for the path of the successive bipartitioning.
Consider the $\gamma=G_1|G_2|\dots|G_{\abs{\gamma}}$ clustering based on the ``connectivity''
with respect to the two-orbital correlations \eqref{eqr:twoorbitMutInf}.
That is, 
the parts $G\in\gamma$ are the sets of orbitals
being the vertices of the connected components of
the \emph{two-orbital correlation graph.}\cite{Rissler-2006,Barcza-2015}
(This is the graph with vertices being the orbitals $i\in M$
and with edges of weights $C_{i|j}(\varrho_{\{i,j\}})$ above a threshold $T_\text{b}$.
We call this \emph{two-orbital correlation clustering.})
It is proven to be a good strategy to do the successive bipartitioning
with respect to $\gamma$, that is, not to split apart the parts of $\gamma$.
Following this strategy, one reaches $\gamma$;
$C_\xi$ increases rapidly after $\gamma$,
but it is not sure that $C_\xi$ increases slowly before $\gamma$.
This is because of the possibility of the
existence of \emph{hidden correlations},
which is an interesting feature of the multiorbital setting.
For example, there exist states in which all two-orbital correlations \eqref{eqr:twoorbitMutInf} are zero,
but the states are correlated as a whole, that is, they cannot be written in a nontrivial product form 
(see the Appendix).
This means that if we follow the above strategy, then $C_\xi$ may change rapidly before we reach $\gamma$.
In this case, $\beta$ does not equal to $\gamma$, but coarser than $\gamma$.

We have investigated the two-orbital and the multiorbital correlation clustering
for the aforementioned molecules.

The two-orbital correlations are drawn by different shades of grey lines in
subfigures (a) of Fig.~\ref{fig:resultscyclic}, \ref{fig:resultscyclic2} and \ref{fig:resultsC2H2x}.
The two-orbital correlation clusterings $\gamma$ are based on the appropriate threshold values $T_\text{b}$.
The distributions of two-orbital correlations, 
and the possible two-orbital correlation thresholds $T_\text{b}$ leading to the known bond structure in the given cases
are shown in subfigures (b). 
For $\mathrm{C}_2\mathrm{H}_{2}$, there is a much wider range for $T_\text{b}'$, leading to triple bond in $\gamma'$,
than for $T_\text{b}$, leading to double bond in $\gamma$,
and
for $\mathrm{C}_2$, there is a much wider range for $T_\text{b}'$, leading to quadruple bond in $\gamma'$,
than for $T_\text{b}$, leading to triple bond in $\gamma$.
A drawback of the two-orbital correlation clustering method is that,
although the two-orbital correlation \eqref{eqr:twoorbitMutInf} is bounded by $0\leq C_{i|j}\leq 2(\ln4)$ uniformly,
a uniform threshold covering all the cases is contained in a quite narrow range $0.269(\ln4)<T_\text{b}\leq0.307(\ln4)$.
The reason for this is that
an orbital seems to be \emph{forced to share its (two-orbital-)correlations}
among the ones strongly correlated with it,
which may be a manifestation of the \emph{monogamy of entanglement\cite{Koashi-2004} in correlations.}
(Different thresholds for the different cases may be obtained 
based on the separation of the correlation scales,
however, this leads to a bond-interpretation rather arbitrary.)

The multiorbital correlation clustering $\beta$,
determined by the use of the method described above,
is drawn by solid red lines in
subfigures (a) of Fig.~\ref{fig:resultscyclic}, \ref{fig:resultscyclic2} and \ref{fig:resultsC2H2x}.
The values of $C_\xi$ during the successive bipartitioning are shown in subfigures (c).
In the cases of the cyclic molecules (Fig.~\ref{fig:resultscyclic} and \ref{fig:resultscyclic2})
we could find a well-defined $\beta$ bond split,
after which the value of $C_\xi$ jumps about at least twice as large
as the maximal step before that.
Note that in certain positions, some humps are observed in the tendencies $C_\xi$
(designated with red rectangles).
These are the effects of correlated clusters of size more than two:
When the successive bipartitioning reaches such a cluster,
following the first large step, smaller steps become possible,
leading to this concave behaviour.
Such humps are coming from two origins.
 The more characteristic one is the cluster of the inner bonding $2\mathrm{p}_\mathrm{z}$ orbitals (denoted with $X_\text{in}$ in the figures).
In the cases of \emph{borole} and \emph{cyclobutadiene},
these humps can be found directly before $\beta$,
they are not steep enough to keep these orbitals together during the successive bipartitioning,
contrary to those in the cases of  \emph{benzene}, \emph{pyrrole}, \emph{furan} and \emph{thiophene},
when these humps can be found after $\beta$.
In this way we can distinguish between aromaticity and antiaromaticity.
 On the other hand,
in the cases of \emph{furan} and \emph{thiophene} (Fig.~\ref{fig:resultscyclic2}),
there is an additional weaker multiorbital correlated structure in each case
besides the aromatic rings,
due to the hyperconjugative interaction of the lone pair with the adjacent $\sigma$-bonds\cite{Szilvasi-2015} (denoted with $X_\text{hc}$ in the figures).
The correlations in them are not strong enough to keep these orbitals together during the successive bipartitioning.
(The almost uniform increase after $\beta$ comes from the bipartitioning of the two-orbital clusters.)
In the cases of the $\mathrm{C}_2\mathrm{H}_{2x}$ sequence (Fig.~\ref{fig:resultsC2H2x}),
it can be seen how 
the correlation picture becomes more and more fuzzy.
Interestingly, for the case of $\mathrm{C}_2\mathrm{H}_2$, 
investigating the tendency $C_\xi$ during the successive bipartitioning in subfigure (c),
one can see that it changes significantly
at the partition leading to a double bond $\beta$ ($=\gamma$),
and there is a much less significant change
at the partition leading to a triple bond $\beta'$ ($=\gamma'$).
For bipartite correlation clustering, $\gamma'$ were more plausible than $\gamma$,
however, here $\beta$ seems to be more plausible than $\beta'$.
This is indeed a very interesting observation, 
which might be to some extent an indication of hidden correlation 
($\beta$ is coarser than $\gamma'$). 
Note however, that despite not being divided in two 
in our multiorbital correlation point of view, 
the four-orbital bond does not contradict the classical picture of triple-bonded $\mathrm{C}_2\mathrm{H}_2$, 
as it contains four electrons.
For the case of $\mathrm{C}_2$, 
investigating the tendencies $C_\xi$ during the successive bipartitioning in subfigure (c),
one cannot give a well-defined bond split $\beta$ besides the $1+8+1$-orbital partition,
because of the high multiorbital correlation of the eight orbitals.
(The splits $\gamma$ and $\gamma'$, given by the two-orbital correlation clustering,
are drawn by solid orange lines in subfigure (a).
In subfigure (c), we show two different paths of the successive refinement 
in the partitioning of the eight bonding orbitals,
the upper one shows a more significant hump,
while the lower one leads through the triple bond $\gamma$.)
That is, 
according to our observations,
there exists no well-defined multiorbital correlation clustering.
The bonding situation in the multireference $\mathrm{C}_2$ is well known as a long-standing puzzle, 
and several bond orders have been suggested, 
including the extensively debated quadruple bond.\cite{Shaik-2012,Shaik-2013, Grunenberg-2012, Frenking-2012, Zhong-2016}
In spite of the fact that four strong two-orbital correlations have been found,
from the reason mentioned above, 
we cannot give a decisive multiorbital correlation answer on the bond order of $\mathrm{C}_2$.

%%%%%%%%%%%%%%%%%%%%%%%%%%%%%%%%%%%%%%%%%%%%%%%%%%%%%%%%%%%%%%%%%%%%%%%%%%%%%%%%
%%%%%%%%%%%%%%%%%%%%%%%%%%%%%%%%%%%%%%%%%%%%%%%%%%%%%%%%%%%%%%%%%%%%%%%%%%%%%%%%
\subsection*{Application 2: Bonds, formed by orbitals}

The bonds, that is, the highly correlated clusters, given by the parts $B\in\beta$, are identified in the previous point.
Now, we can investigate the correlations inside the bonds $B\in\beta$.
For this purpose, we 
use the $k$-partitionability and $k$-producibility correlation
$C_{\text{$k$-part},B}$ and $C_{\text{$k$-prod},B}$, respectively,
(see \eqref{eqr:kppCorr}),
both of them are considered with respect to the splits $\Pi(B)$.

The results are again summarised in Fig.~\ref{fig:resultscyclic}, \ref{fig:resultscyclic2} and \ref{fig:resultsC2H2x}.
For the two-orbital bonds $B=\{i,j\}\in\beta$, the important quantities
boil down to the two-orbital correlation \eqref{eqr:twoorbitMutInf},
$C_{\text{$2$-part},\{i,j\}}=C_{\text{$1$-prod},\{i,j\}}=C_{i|j}$.
Its magnitude can be read off from subfigures (b).
More interesting is the case of bonds consisting of more than two orbitals.
The $2\mathrm{p}_\mathrm{z}$ orbitals (contained in $X_\text{in}\subset M$)
in the cases of \emph{benzene}, \emph{pyrrole}, \emph{furan} and \emph{thiophene} form aromatic bond, and
in the cases of \emph{borole} and \emph{cyclobutadiene} do not.
This can be seen in 
the full increasing and decreasing tendencies
$C_{\text{$k$-part},X_\text{in}}$ and
$C_{\text{$k$-prod},X_\text{in}}$, shown in subfigures (d):
on the one hand, $C_{\text{$2$-part},X_\text{in}}= C_{\text{$(\abs{X_\text{in}}-1)$-prod},X_\text{in}}$ is high in the four aromatic cases, 
that is, the orbitals in $X_\text{in}$ cannot be split even into two parts,
and, accordingly, the greatest part is of size $\abs{X_\text{in}}$;
on the other hand, $C_{\text{$2$-prod},X_\text{in}}$ and 
$C_{\text{$2$-part},X_\text{in}}$, respectively,
$C_{\text{$3$-part},X_\text{in}}$ 
are low in the two antiaromatic cases,
that is, the orbitals in $X_\text{in}$ can be split into parts of size at most $2$,
their number are $2$, respectively $3$.
So we can distinguish between aromaticity and antiaromaticity also in this way.
The orbitals participated in the hyperconjugative interaction in \emph{furan} and \emph{thiophene}
(contained in $X_\text{hc}\subset M$)
show weaker correlation than the ones in the aromatic ring.
In the cases of the $\mathrm{C}_2\mathrm{H}_{2x}$ sequence,
the orbitals participating in the bonds between the carbon atoms
(contained in $X_\text{in}$)
are getting more and more multiorbital-correlated.
For $\mathrm{C}_2\mathrm{H}_{4}$, 
$C_{\text{$2$-part},X_\text{in}}$ and $C_{\text{$2$-prod},X_\text{in}}$ are near zero, 
that is, the two two-orbital bonds can be considered independent.
For $\mathrm{C}_2\mathrm{H}_{2}$,
$C_{\text{$2$-part},X_\text{in}}$ and $C_{\text{$4$-prod},X_\text{in}}$ are negligibly low,
 while $C_{\text{$3$-part},X_\text{in}}$ is significant,
leading again to a double-bond, containing a four-orbital one.
For $\mathrm{C}_2$,
$C_{\text{$2$-part},X_\text{in}}$, although being relatively low, does not seem to be completely negligible.
In the latter two cases, we can see now from a local point of view,
which was proposed in the previous Section from a global point of view,
that $X_\text{in}$ is not divided \emph{completely} into independent bonds.

%%%%%%%%%%%%%%%%%%%%%%%%%%%%%%%%%%%%%%%%%%%%%%%%%%%%%%%%%%%%%%%%%%%%%%%%%%%%%%%%
%%%%%%%%%%%%%%%%%%%%%%%%%%%%%%%%%%%%%%%%%%%%%%%%%%%%%%%%%%%%%%%%%%%%%%%%%%%%%%%%
\subsection*{Application 3: Molecule, formed by atoms}
An \emph{atom} is now represented by an $A\subseteq M$ set of orbitals,
where the orbitals $i\in A$ are the ones localised on the given atom.
The molecule can be considered as a set of atoms,
this can be represented by the split $\alpha=A_1|A_2|\dots|A_{\abs{\alpha}}\in\Pi(M)$ (\emph{atomic split})
of the molecule.
Here an important quantity is the $\alpha$-correlation $C_\alpha(\varrho_M)$,
 and the $\alpha$-coarsened
 $k$-partitionability and $k$-producibility correlations
 $C_{\text{$k$-part},\alpha}(\varrho_M)$, $C_{\text{$k$-prod},\alpha}(\varrho_M)$.
These characterise the different aspects of the correlations with respect to the atomic split $\alpha$.

The atomic split $\alpha$
for the aforementioned molecules
are drawn by dashed blue lines in subfigures (a) of Fig.~\ref{fig:resultscyclic}, \ref{fig:resultscyclic2} and \ref{fig:resultsC2H2x}.
The values of $C_\alpha$ are also shown.
Calculating $C_{\text{$k$-part},\alpha}(\varrho_M)$ and 
$C_{\text{$k$-prod},\alpha}(\varrho_M)$ is infeasible,
due to the large density matrices of high entropy,
however, note that we already have the largest members of these hierarchies, since 
$C_{\text{$\abs{\alpha}$-part},\alpha}(\varrho_M)= C_{\text{$1$-prod},\alpha}(\varrho_M)
= C_\alpha(\varrho_M)$.
The value of this is near $C_\bot(\varrho_M)$,
that is, as can be expected, the atoms are strongly correlated with one another in the molecules.

%%%%%%%%%%%%%%%%%%%%%%%%%%%%%%%%%%%%%%%%%%%%%%%%%%%%%%%%%%%%%%%%%%%%%%%%%%%%%%%%
%%%%%%%%%%%%%%%%%%%%%%%%%%%%%%%%%%%%%%%%%%%%%%%%%%%%%%%%%%%%%%%%%%%%%%%%%%%%%%%%
\subsection*{Application 4: Atoms, formed by orbitals}
The orbitals localised on given atoms are collected in the parts $A\in\alpha$ in the previous point.
Now, we can investigate the correlations in the atoms $A\in\alpha$.
For this purpose, we  
use the $k$-partitionability and $k$-producibility correlation
$C_{\text{$k$-part},A}$ and $C_{\text{$k$-prod},A}$, respectively,
(see \eqref{eqr:kppCorr}),
both of them are considered with respect to the splits $\Pi(A)$.

We have investigated the nontrivial (non-$\mathrm{H}$) atoms 
in the aforementioned molecules.
Although not all the positions of the $\mathrm{C}$ atoms are equivalent 
in the molecules, 
the correlation measures take roughly the same values for those.
The full increasing and decreasing tendencies
$C_{\text{$k$-part},A}(\varrho_A)$ and
$C_{\text{$k$-prod},A}(\varrho_A)$ are shown in subfigures (e).
Note that the values of these are usually smaller 
than the correlations in the bonds, by about two orders of magnitude.
In the sequence $\mathrm{C}_2\mathrm{H}_4$, $\mathrm{C}_2\mathrm{H}_2$ and $\mathrm{C}_2$
we can also see, how the increase of the multiorbital correlations leads to
more and more strong correlations among the orbitals localised on the same $\mathrm{C}$ atom.
The hyperconjugative interaction leads to the same results on the $\mathrm{O}$ and $\mathrm{S}$ atoms in furan and thiophene.

%%%%%%%%%%%%%%%%%%%%%%%%%%%%%%%%%%%%%%%%%%%%%%%%%%%%%%%%%%%%%%%%%%%%%%%%%%%%%%%%
%%%%%%%%%%%%%%%%%%%%%%%%%%%%%%%%%%%%%%%%%%%%%%%%%%%%%%%%%%%%%%%%%%%%%%%%%%%%%%%%
\subsection*{Remarks on the applications}
Having the results of all the four applications in hand,
we can now observe how the sum rule \eqref{eqr:totalCorrpart} works.
In the first two applications, when we considered the $\beta$ bond split, 
$\sum_{B\in\beta} C_{\bot,B}(\varrho_B)$ was large and $C_\beta(\varrho_M)$ small;
while 
in the second two applications, when we considered the $\alpha$ atomic split, 
$\sum_{A\in\alpha} C_{\bot,A}(\varrho_A)$ was small and $C_\alpha(\varrho_M)$ large,
and these are connected by the sum rule \eqref{eqr:totalCorrpart} as
\begin{equation}
\label{eqr:abtotalCorrpart}
\begin{split}
C_\bot(\varrho_M)
&= \sum_{A\in\alpha} C_{\bot,A}(\varrho_A) + C_{\alpha}(\varrho_M)\\
&= \sum_{B\in\beta } C_{\bot,B}(\varrho_B) + C_{\beta }(\varrho_M).
\end{split}
\end{equation}
Based on these, we can consider the molecule as 
the weakly correlated set of strongly correlated bonds, or
the strongly correlated set of weakly correlated atoms.
Note that this holds for the equilibrium structure, which is the only one considered here.
If the internuclear distances are altered,
which is a method for the investigation of bond-formation,\cite{Boguslawski-2013,Mottet-2014,Fertitta-2014,Duperrouzel-2014,Zhao-2015}
we expect that the above picture is altered accordingly,
however, the sum rule \eqref{eqr:abtotalCorrpart} holds with altered numerical values.

On the other hand,
we may give a definition of the molecule from a correlation point of view:
the orbitals $M$ form a molecule, 
if there exists no nontrivial partition 
which is coarser than $\alpha$, describing the atoms,
and $\beta$, describing the bonds,
that is, $\alpha\vee\beta=\top$.
(In these cases, intermolecular bonds do not appear in $\beta$.)

%%%%%%%%%%%%%%%%%%%%%%%%%%%%%%%%%%%%%%%%%%%%%%%%%%%%%%%%%%%%%%%%%%%%%%%%%%%%%%%%
%%%%%%%%%%%%%%%%%%%%%%%%%%%%%%%%%%%%%%%%%%%%%%%%%%%%%%%%%%%%%%%%%%%%%%%%%%%%%%%%
\section*{Conclusions and outlook}

We have presented a novel theory of the chemical bond 
which is inspired by quantum information theory and based on multiorbital correlations.
Contrary to the literature,
where only two-orbital notions were considered,
we have invented and used true multiorbital notions.
Illustrating the use of this theoretical toolbox,
we have investigated several small prototypical molecules
and showed how in a black-box manner the bonding picture of a molecule 
naturally comes out from the multiorbital correlations of occupations of localised atomic-like orbitals.
We have identified
the bonds with the strongly correlated clusters,
and
characterised quantitatively how well a given bonding picture describes these molecules.
Our tools are, e.g.,
able to distinguish between aromaticity and antiaromaticity in cyclic conjugated systems.
On the example of the sequence $\mathrm{C}_2\mathrm{H}_{2}$, $\mathrm{C}_2\mathrm{H}_{4}$, and $\mathrm{C}_2\mathrm{H}_{6}$,
we have seen that the increase of wide-range multiorbital correlations results
in the decrease of the well-posedness of multiorbital correlation clustering.
In the extreme case of $\mathrm{C}_2$, this leads to
the nonexistence of a well-defined multiorbital correlation clustering,
which provides a reason for the debated bonding picture.

We would like to emphasise again that
the treatment in terms of true multiorbital correlations
seems to be a very natural point of view
in the investigation of bonding among more than two atoms.
The multiorbital correlation based quantities
have their statistical meaning on their own right,
and we have already seen several of their applications.
However, it would be interesting to relate them
to other standard quantities in quantum chemistry,
quantifying, e.g., bond strength or aromaticity.
Besides aromaticity,
this treatment may find applications also in
 multicenter transition metal cluster chemistry.

We have seen how the multiorbital correlations
characterise the chemical bonds, if the orbitals are localised.
We note, however, that the theory can also be applied to any (orthonormalized) sets of orbitals,
then it characterises the correlations among those orbitals.
These, of course, do not have to be related to the chemical bonds,
but may be related to other chemical properties.

From the point of view of theoretical power and beauty,
the multiorbital correlation theory provides a much more natural and flexible treatment for multiorbital situations
than using only two-orbital correlations, done in preceding works.
An example supporting this is given by the \eqref{eqr:abtotalCorrpart} application of the sum rule \eqref{eqr:totalCorrpart}.
Contrary to this, a treatment based only on two-orbital correlations
is theoretically hard to grasp, due to monogamy-like issues of correlations in quantum systems.
This is why the notion of hidden correlations is not well-defined,
and 
to formulate a clear-cut (quantitative and/or operative) definition
is an open question.

%%%%%%%%%%%%%%%%%%%%%%%%%%%%%%%%%%%%%%%%%%%%%%%%%%%%%%%%%%%%%%%%%%%%%%%%%%%%%%%%
\section*{Methods}
For the numerical results shown in this paper
we have performed calculations using the 
quantum chemistry version of the \emph{density matrix renormalization group} (QC-DMRG) algorithm.%
\cite{White-1999,White-1992b,Marti-2008b,Zgid-2008,Kurashige-2009,Legeza-2014,Wouters-2014e,Szalay-2015a,Chan-2015,Olivares-2015}
We have controlled the numerical accuracy using the \emph{dynamic block state selection} (DBSS) approach\cite{Legeza-2003a,Legeza-2004b,Szalay-2015a}
and the maximum number of block states varied in the range of 500-4000 
for an a priory set quantum information loss threshold value $\chi=10^{-6}$.
The ordering of molecular orbitals along the one-dimensional topology of the DMRG
was optimised using the Fiedler approach\cite{Barcza-2011,Fertitta-2014}
and the active space was extended dynamically 
based on the \emph{dynamically extended active space} (DEAS) procedure.\cite{Legeza-2003b,Szalay-2015a}
We have used DMRG to obtain the optimised MPS wavefunction,
which was then used to construct the reduced density matrices,
from which the correlation measures \eqref{eqr:xiCorr} and \eqref{eqr:kppCorr} were calculated.

Geometries have been optimised at HF/cc-pVTZ level of theory
which yielded sufficient geometries in accordance with higher level methods.
To obtain the localised atomic orbitals for the DMRG procedure,
we first optimised the exponents of the STO-6G basis set using the MRCC program\cite{MRCC,Rolik-2013,Mester-2015}
which approach resulted in similar HF energy to the cc-pVTZ basis set result within $10^{-2}$ Hartree.
Then we used the Pipek-Mezey procedure\cite{Pipek-1989}
implemented in MOLPRO\cite{MOLPRO} Version 2010.1,
with tight threshold  $10^{-12}$, and minimised the number of atomic orbitals contributed to each localised orbitals.
All localised orbitals have been used in the DMRG procedure
thus, as a result, we have carried out calculations at the FCI limit for all molecules.
Then the results close to the FCI limit have been analysed in the paper.\cite{Szilvasi-2015}
We note that we also calculated all results using HF/STO-3G geometry and localised STO-3G orbitals,
with literature value exponents,
and found neglectable difference compared to the results presented in the manuscript.
(These results are presented in the Appendix for comparison.)
This suggests that our analysis is very robust in general.
We note that this robustness is not entirely surprising.
Mayer has shown\cite{Mayer-1983,Mayer-2007,Mayer-2016} that
extracting chemical information from molecular wavefunctions
such as bond orders and valence indices could also be obtained using only STO-3G basis set.

The datasets generated during and/or analysed during the current study
are available from the corresponding author on reasonable request.

%%%%%%%%%%%%%%%%%%%%%%%%%%%%%%%%%%%%%%%%%%%%%%%%%%%%%%%%%%%%%%%%%%%%%%%%%%%%%%%%
% Addendum

\section*{Additional information}
\textbf{Acknowledgements:}
Discussions with \emph{P\'eter Vrana} and \emph{Zolt\'an Zimbor\'as} 
are gratefully acknowledged.
Sz.Sz., G.B.~and {\"O}.L.~are supported by 
the \textit{Hungarian Scientific Research Fund} (project ID: OTKA-NN110360),
the \textit{National Research, Development and Innovation Office} (project ID: NKFIH-K120569)
and 
the \textit{``Lend\"ulet''} program of the Hungarian Academy of Sciences.
L.V.~is supported by 
the \emph{Grant Agency of the Czech Republic} (grant no.~16-12052S).

\textbf{Author contributions statement:}
All authors took part in the analysis of the data 
and contributed to the discussion about the results 
and reviewed the manuscript.
Sz.Sz.~invented the theoretical background in multipartite correlations,
proposed the multipartite correlation measures
and formulated the clustering method.
T.Sz.~and L.V.~conceived and designed the investigation. 
G.B.~and {\"O}.L.~developed the computer implementations.
T.Sz., L.V.~and G.B.~performed the calculations.

\textbf{Correspondence:}
Correspondence and requests for materials
about multiorbital correlation theory 
should be addressed to Sz.Sz.~(e-mail: \href{mailto:szalay.szilard@wigner.mta.hu}{\texttt{szalay.szilard@wigner.mta.hu}}),
about the theory of chemical bonds 
should be addressed to T.Sz.~(e-mail: \href{mailto:szilvasitibor@ch.bme.hu}{\texttt{szilvasitibor@ch.bme.hu}}),
about the use of the DMRG method
should be addressed to {\"O}.L.~(e-mail: \href{mailto:legeza.ors@wigner.mta.hu}{\texttt{legeza.ors@wigner.mta.hu}}).

\textbf{Competing financial interests:}
The authors declare no competing financial interests.

%%%%%%%%%%%%%%%%%%%%%%%%%%%%%%%%%%%%%%%%%%%%%%%%%%%%%%%%%%%%%%%%%%%%%%%%%%%%%%%%
\bibliographystyle{apsrev4-1}
\bibliography{paper_correlbond}{}

\appendix
%%%%%%%%%%%%%%%%%%%%%%%%%%%%%%%%%%%%%%%%%%%%%%%%%%%%%%%%%%%%%%%%%%%%%%%%%%%%%%%%
\section{Multipartite correlations}

Here we recall the two-level theory of multipartite correlations,
introduced earlier for the investigation of multipartite entanglement\cite{Szalay-2015b}.
Furthermore, we extend the construction with the formalism of restriction to subsystems and coarsening,
and we also construct bounds and relations for the resulting correlation measures.
Using these tools we formulate and solve the task
of multipartite correlation clustering,
that is, dividing the whole system
into weakly correlated subsystems consisting of strongly correlated elementary subsystems.

\begin{table*}
\setlength{\tabcolsep}{6pt}
\begin{tabular}{p{3.4cm}|p{4.2cm}|p{4.2cm}|p{4.2cm}}
quantum information 
\newline theory & 
quantum mechanics 
\newline (distinguishable particles)
\newline (first quantized) & 
many-body quantum physics
\newline (second quantized) & 
quantum chemistry
\newline (Born-Oppenheimer appr.)
\newline (second quantized)\\
\hline
system & ensemble of particles & chain, lattice & electronic system of molecule \\
elementary subsystem & particle & site & orbital \\
composite subsystem & subensemble of particles & block/cluster (of sites) & cluster (of orbitals) \\
\end{tabular}
\caption{Dictionary. 
In the main text we follow mainly the quantum chemistry language, because of the illustrative chemical applications,
while in this Appendix we use the general quantum information theory language.
Note that the elementary subsystems are always distinguishable in this treatment.
We do not treat the correlations in the first quantized formalism of indistinguishable particles.
}
\label{tab:dict}
\end{table*}

%%%%%%%%%%%%%%%%%%%%%%%%%%%%%%%%%%%%%%%%%%%%%%%%%%%%%%%%%%%%%%%%%%%%%%%%%%%%%%%%
\subsection{Setting the stage.}

Consider a quantum system composed of $n$ elementary subsystems, labelled by natural numbers $L=\set{1,2,\dots,n}$.
For the quantum mechanical description of every elementary subsystem, a Hilbert space of uniform dimension $d$ is used.
These can form the whole system investigated, or only a subsystem of that.
Note that, in the Appendi, we give a general treatment.\cite{Ohya-1993,Nielsen-2000,Wilde-2013,Petz-2008}
In the case in which we use these tools in the main text,
the elementary subsystems are the orbitals (or clusters of orbitals in coarsened cases),
that is, we use the second quantized formalism there,
contrary to the first quantized formalism, in which case the elementary subsystems are the electrons.
(A dictionary is given in Table \ref{tab:dict}.)
Note that, however, this general treatment works equally well in the first and the second quantized pictures,
this is why we have chosen this way of presentation here.
In the first quantized picture, 
the construction characterises the correlations of the, 
e.g., position (or spin or other degrees of freedom) of different distinguishable particles,
and in the second quantized picture,
the same construction characterises the correlations of the 
occupation of states with different, e.g., position (or spin or other degrees of freedom).
We also note that in the case of fermionic particles,
the occupations of the sites are given in terms of anticommuting operators.
This leads to some difficulties when one uses tools working well in the distinguishable case.
However, it has turned out that
all what we need during the construction of our multipartite/multiorbital correlation clustering
are working well if the situation is restricted to
the physical subspace of the operator algebra,\cite{Araki-2003b,Moriya-2005,Banuls-2007}
consisting of parity-preserving operators.
Since in the molecular-physical situations, considered in the main text,
even the preservation of the particle number holds,
the following construction can obviously be applied.

The \emph{state} of the quantum system is represented by
a normalised positive linear functional acting on the algebra of the observables.
It is given by the density operator $\varrho_L$,
which is a positive semidefinite operator of trace one,
acting on the Hilbert space associated to the system.\cite{Ohya-1993,Wilde-2013,Petz-2008}
By restricting the state to the subalgebra of a (not necessarily elementary) subsystem $X\subseteq L$,
one can form the reduced density operator $\varrho_X$ of the subsystem.
An essential property of a quantum state is its \emph{mixedness}.
It can be characterised by %, among others, 
the von Neumann entropy\cite{Neumann-1927,Ohya-1993,Petz-2008}
\begin{equation}
\label{eq:vonNeumann}
S(\varrho) = -\tr \varrho \ln \varrho.
\end{equation}
One can also compare two quantum states %, among others,
in the sense of statistical distinguishability%
\cite{Hiai-1991}
by the Umegaki relative entropy\cite{Umegaki-1962,Ohya-1993,Sagawa-2012}
(or quantum Kullback-Leibler divergence)
\begin{equation}
\label{eq:Umegaki}
D(\varrho||\omega) = \tr \varrho (\ln \varrho - \ln \omega).
\end{equation}
These two functions are of central importance 
in quantum information theory,\cite{Nielsen-2000,Wilde-2013,Petz-2008}
and the whole construction we build here is based on them.
Both of these functions are nonnegative,\cite{Umegaki-1962,Ohya-1993}
and they have several beautiful properties, making them extremely useful,
also in the cases in which we apply them in the sequel.
Maybe the most important one is the \emph{strong subadditivity}
of the von Neumann entropy\cite{Lieb-1973,Ohya-1993,Araki-2003b}
\begin{subequations}
\begin{equation}
\label{eq:vonNeumannSSA}
S(\varrho_{X\cup X'}) + S(\varrho_{X\cap X'}) \leq S(\varrho_{X}) + S(\varrho_{X'}).
\end{equation}
A special case for disjoint subsystems $X$ and $X'$ is the \emph{subadditivity}
of the von Neumann entropy, 
\begin{equation}
\label{eq:vonNeumannSA}
S(\varrho_{X\cup X'}) \leq S(\varrho_{X}) + S(\varrho_{X'}).
\end{equation}
From the strong subadditivity,
the so called \emph{Araki-Lieb triangle inequality}
of the von Neumann entropy\cite{Araki-1970,Moriya-2005,Ohya-1993,Petz-2008,Wilde-2013} also follows
\begin{equation}
\label{eq:vonNeumannTriangle}
\bigabs{S(\varrho_{X}) - S(\varrho_{X'})}
\leq S(\varrho_{X\cup X'}). 
\end{equation}
\end{subequations}
Maybe even more fundamental is the \emph{monotonicity}
of the relative entropy
\emph{with respect to state reduction,}\cite{Petz-1986,Ohya-1993,Petz-2003,Kosaki-1986,Araki-2003b}
that is, for subsystems $Y\subseteq X$,
\begin{equation}
\label{eq:UmegakiRedMon}
D(\varrho_X||\omega_X) \geq D(\varrho_Y||\omega_Y).
\end{equation}

%%%%%%%%%%%%%%%%%%%%%%%%%%%%%%%%%%%%%%%%%%%%%%%%%%%%%%%%%%%%%%%%%%%%%%%%%%%%%%%%
\subsection{Level I correlations.}

\begin{figure*}
\includegraphics{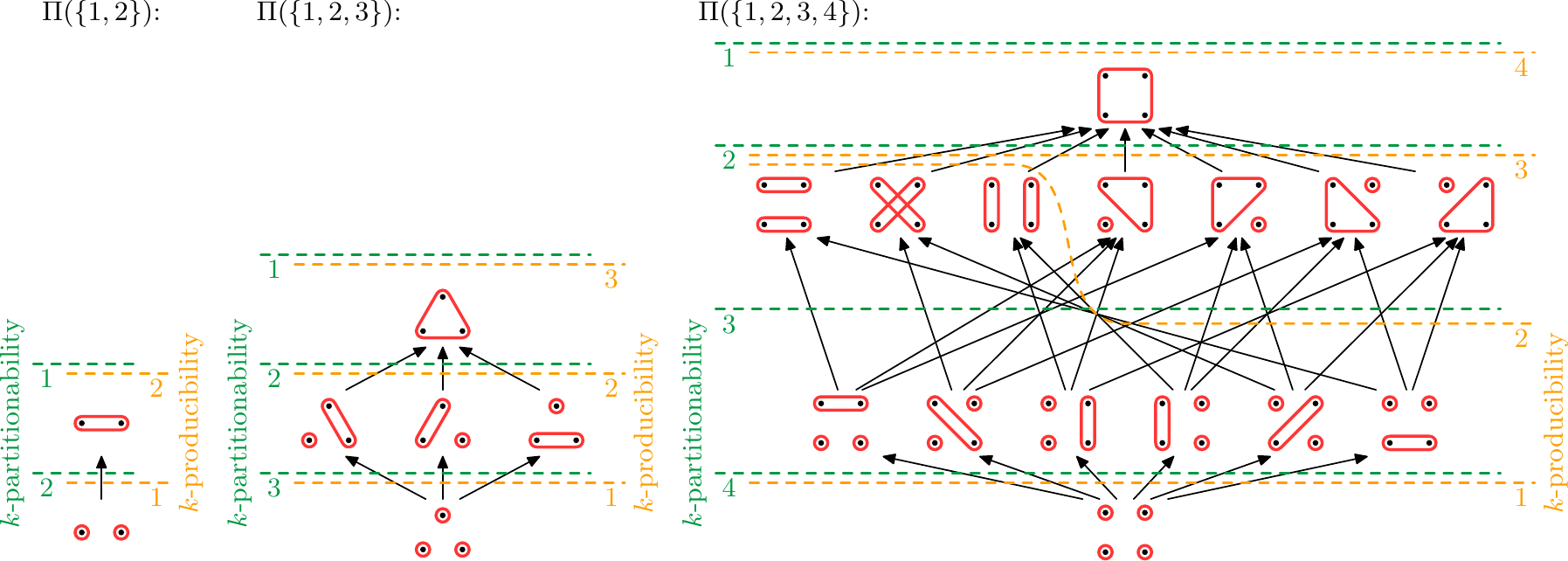}
\caption{Lattices of partitions for $n=2,3,4$ subsystems.
The covering is depicted with arrows from the finer to the coarser partition.
The nonempty down-sets corresponding to the notions of $k$-partitionably uncorrelated
and $k$-producibly uncorrelated
are located under the dashed lines starting from the left and the right, respectively. 
}
\label{fig:partlattices}
\end{figure*}

A split of the system into \emph{parts} is given by the \emph{partition} 
$\xi =\set{X_1,X_2,\dots,X_{\abs{\xi}} } \equiv X_1|X_2|\dots|X_{\abs{\xi}}$
of the labels $L$, that is, the $X\in\xi$ sets of labels are nonempty and disjoint,
and their union is the whole $L$.
The set of the partitions of $L$ is denoted with $\Pi(L)$.
The partitions are illustrated with small pictographs in Fig.~\ref{fig:partlattices} for $n=2,3,4$ subsystems.
For two partitions, % $\upsilon,\xi$, 
$\upsilon$ is a \emph{refinement}\cite{Davey-2002} of $\xi$,
(``$\upsilon$ is \emph{finer} than $\xi$'', or, ``$\xi$ is \emph{coarser} than $\upsilon$'')
denoted with $\upsilon\finereq\xi$,
if $\xi$ can be obtained from $\upsilon$ 
by joining some (maybe none) of the parts of $\upsilon$.
That is,
\begin{equation}
\label{eq:oderingParts}
\upsilon\finereq\xi\quad\overset{\text{def.}}{\Longleftrightarrow}\quad
\forall Y\in\upsilon, \exists X\in\xi:\; Y\subseteq X.
\end{equation}
The refinement relation is a partial order, 
and the set of partitions $\Pi(L)$ turns out to be a lattice\cite{Davey-2002}.
The top and bottom elements are 
$\top=\{\{1,2,\dots,n\}\}=\{L\}$ and
$\bot=\{\{1\},\{2\},\dots,\{n\}\}\equiv1|2|\dots|n$, respectively.
Later we will also need the notion of being neighbours in the lattice.
This is called covering relation.\cite{Davey-2002}
For two partitions,
$\xi$ \emph{covers} $\upsilon$, denoted with $\upsilon\covered\xi$, if
$\abs{\upsilon}=\abs{\xi}+1$, while $\upsilon\finer\xi$,
from which we can conclude that 
there exists exactly one part $X_*\in\xi$,
for which there are exactly two parts $Y_{*1},Y_{*2}\in\upsilon$,
such that $X_*=Y_{*1}\cup Y_{*2}$;
and all the other parts in $\xi$ can also be found in $\upsilon$.
That is, 
\begin{equation}
\label{eq:coveringParts}
\upsilon\covered\xi\quad\Longleftrightarrow\quad
\left\{
\begin{aligned}
\upsilon\setminus\xi &= \{Y_{*1},Y_{*2}\}\in\Pi(X_*),\quad\text{and}\\
\xi\setminus\upsilon &= \{X_*\}\in\Pi(X_*).
\end{aligned}
\right.
\end{equation}
The covering relation is illustrated with arrows in Fig.~\ref{fig:partlattices}
for $n=2,3,4$ subsystems.

A quantum state $\varrho_L$ is \emph{$\xi$-uncorrelated},
that is, uncorrelated with respect to the partition $\xi$,
if it can be written in a product form of the reduced states with respect to $\xi$.
One can characterise 
to what extent a state is not $\xi$-uncorrelated.
To this end, let the \emph{$\xi$-correlation}\cite{Szalay-2015b},
also called ``among-the-clusters correlation information''\cite{Herbut-2004} be
defined as
\begin{equation}
\label{eq:xiCorr}
C_\xi(\varrho_L) := \sum_{X\in\xi} S(\varrho_X) - S(\varrho_L),
\end{equation}
with the reduced states $\varrho_X$.
(For the finest $\bot$ split, this is also called ``correlation information''\cite{Herbut-2004},
or ``multipartite mutual information''\cite{Yang-2009},
or ``total correlation''\cite{Legeza-2004b,Legeza-2006b},
also considered by Lindblad\cite{Lindblad-1973} and used\cite{Horodecki-1994} to describe correlations within
multipartite quantum systems.)
For a bipartition $\xi=X_1|X_2\equiv X_1|(L\setminus X_1)$, 
the quantity $C_{X_1|X_2}(\varrho_L) = S(\varrho_{X_1}) + S(\varrho_{X_2}) - S(\varrho_L) = I_{X_1|X_2}(\varrho_L)$
is the well-known (bipartite) mutual information.\cite{Ohya-1993,Petz-2008,Nielsen-2000,Wilde-2013}
The information-geometrical meaning of the $\xi$-correlation is also clarified\cite{Modi-2010,Szalay-2015b},
it expresses the minimal distinguishability \eqref{eq:Umegaki} of a state from the set of $\xi$-uncorrelated states,
the ``relative entropy of $\xi$-correlation'',
\begin{equation}
\label{eq:geomCorrImin}
\min_{\omega_L \text{ $\xi$-uncorr.}} 
D\bigl(\varrho_L\big\Vert\omega_L\bigr) =
C_\xi(\varrho_L).
\end{equation}
Note that because of the \eqref{eq:vonNeumannSA} subadditivity of the von Neumann entropy, 
the $\xi$-correlation takes higher value for a finer split,
\begin{equation}
\label{eq:MultiMonI}
\upsilon\finereq\xi\quad\Longleftrightarrow\quad C_\upsilon\geq C_\xi.
\end{equation}
This is called \emph{multipartite monotonicity (of the first kind)}.\cite{Szalay-2015b}

Some remarks on the relation to entanglement\cite{Horodecki-2009,Szalay-2013,Amico-2008} is also on place here.
If the whole system can be described by a pure state $\varrho_L=\cket{\psi_L}\bra{\psi_L}$,
the \eqref{eq:xiCorr} correlation with respect to the $X|(L\setminus X)$ split
is just two times the entanglement entropy\cite{Bennett-1996,Nielsen-2000,Wilde-2013} of subsystem $X\subseteq L$,
that is, $C_{X|(L\setminus X)}(\varrho_L) = 2S(\varrho_X)$,
because of the Schmidt decomposition%
\cite{Schmidt-1907,Nielsen-2000,Wilde-2013}.
Generally speaking,
pure states of classical systems are always uncorrelated.
If a pure state of a quantum system is correlated,
then this correlation is of quantum origin, and it is called \emph{entanglement}.
Then correlation measures for pure quantum states often lead to entanglement measures\cite{Szalay-2015b}.
Mixed states of a classical system can be either correlated or uncorrelated.
If a mixed state of a quantum system is correlated,
this correlation can either be classical or it can contain also quantum entanglement\cite{Modi-2010,Szalay-2011}.
For mixed states, entanglement measures can also be constructed\cite{Szalay-2015b,Szalay-2012}.

%%%%%%%%%%%%%%%%%%%%%%%%%%%%%%%%%%%%%%%%%%%%%%%%%%%%%%%%%%%%%%%%%%%%%%%%%%%%%%%%
\subsection{Level II correlations.}

We also need to consider a second level notion of uncorrelated states.
This expresses that a state is uncorrelated with respect to at least one partition from a given set.
If $\xi$-uncorrelated states are considered,
then $\upsilon$-uncorrelated states are automatically considered also
for all $\upsilon\finereq\xi$,
so it is natural to consider the nonempty \emph{down-sets} of partitions\cite{Szalay-2015b} (also called \emph{order ideal}).
A nonempty $\vs{\xi}=\set{\xi_1,\xi_2,\dots,\xi_{\abs{\vs{\xi}}}}\subseteq\Pi(L)$ set
is a nonempty \emph{down-set}
if it contains every partition which is finer than its maximal elements.\cite{Davey-2002}
This can be drawn by lines cutting the partition lattice into two parts
in a way that the arrows cross these lines in one way only.
The elements of the down-set are then the ones located under that line.
Some of these cuttings are illustrated with dashed lines in Fig.~\ref{fig:partlattices}
for $n=2,3,4$ subsystems.
Here we also have a natural partial order, 
being the standard set-theoretical inclusion among the nonempty down-sets,
$\vs{\upsilon}\finereq\vs{\xi}$ if and only if $\vs{\upsilon}\subseteq\vs{\xi}$,
and the set of nonempty down-sets of partitions is also a lattice\cite{Davey-2002}.
Particular down-sets are the \emph{principal ideals} $\downset\{\xi\}=\sset{\xi'\in\Pi(L)}{\xi'\finereq\xi}$,
being the down-sets of partitions finer than or equal to a single $\xi$.
Easy to check that $\downset\{\xi\}\preceq\downset\{\xi'\}\Leftrightarrow \xi\preceq\xi'$, so the Level I structure is enbedded into Level II in this way.
(A dual way of this embedding is 
using the \emph{principal filters} $\upset\{\xi\}=\sset{\xi'\in\Pi(L)}{\xi\finereq\xi'}$,
which is a particular kind of \emph{up-set} or \emph{order filter}.)

A quantum state $\varrho_L$ is \emph{$\vs{\xi}$-uncorrelated},
if it is $\xi$-uncorrelated with respect to at least one $\xi\in\vs{\xi}$.
One can characterise 
to what extent a state is not $\vs{\xi}$-uncorrelated.
To this end, let the \emph{$\vs{\xi}$-correlation}\cite{Szalay-2015b} be
defined as
\begin{equation}
\label{eq:vsxiCorr}
C_{\vs{\xi}}(\varrho_L) := \min_{\xi\in\vs{\xi}}C_\xi(\varrho_L).
\end{equation}
The information-geometrical meaning of the $\vs{\xi}$-correlation is also clarified\cite{Szalay-2015b},
it expresses the minimal distinguishability \eqref{eq:Umegaki} of a state from the set of $\vs{\xi}$-uncorrelated states,
\begin{equation}
\begin{split}
\label{eq:geomCorrIImin}
\min_{\omega_L \text{ $\vs{\xi}$-uncorr.}} 
D\bigl(\varrho_L\big\Vert\omega_L\bigr)
&=\min_{\forall \xi\in\vs{\xi}} \;
\min_{\omega_L \text{ $\xi$-uncorr.}} 
D\bigl(\varrho_L\big\Vert\omega_L\bigr)\\
&= \min_{\forall \xi\in\vs{\xi}}
C_\xi(\varrho_L)
= C_{\vs{\xi}}(\varrho_L),
\end{split}
\end{equation}
see \eqref{eq:geomCorrImin}.
Note that because of the multipartite monotonicity \eqref{eq:MultiMonI} of the
$\xi$-correlation \eqref{eq:xiCorr},
it is sufficient to calculate the minimum over the maximal elements of $\vs{\xi}$
in the $\vs{\xi}$-correlation \eqref{eq:vsxiCorr}, that is,
$C_{\vs{\xi}}(\varrho_L) = \min_{\xi\in\max\vs{\xi}}C_\xi(\varrho_L)$.
In particular, $C_{\downset\{\xi\}}=C_\xi$ for the principal ideal $\downset\{\xi\}$.
Note on the other hand that, because of the multipartite monotonicity \eqref{eq:MultiMonI} of the
$\xi$-correlation,
the $\vs{\xi}$-correlation takes higher value for a smaller nonempty down-set,
\begin{equation}
\label{eq:MultiMonII}
\vs{\upsilon}\finereq\vs{\xi}\quad\Longleftrightarrow\quad C_{\vs{\upsilon}}\geq C_{\vs{\xi}}.
\end{equation}
This is called \emph{multipartite monotonicity (of the second kind)}\cite{Szalay-2015b}.

%%%%%%%%%%%%%%%%%%%%%%%%%%%%%%%%%%%%%%%%%%%%%%%%%%%%%%%%%%%%%%%%%%%%%%%%%%%%%%%%
\subsection{\texorpdfstring{$k$}{k}-partitionability and \texorpdfstring{$k$}{k}-producibility.}

So far we introduced different kinds of uncorrelated states,
and we characterised to what extent a state is not uncorrelated in the different ways.
This construction actually led to different \emph{notions} of correlations.
For our goals
it is enough to consider some special ones of them.

For $k=1,2,\dots,n$, a state is \emph{$k$-partitionably uncorrelated},
if it can be written in the product form of density matrices of (at least) $k$ subsystems.
That is, they are $\vs{\mu}_k$-uncorrelated for 
the nonempty down-set $\vs{\mu}_k$ containing all the partitions $\mu$
in which the \emph{number ($\abs{\mu}$) of the parts} is greater than or equal to $k$,\cite{Szalay-2015b}
\begin{subequations}
\label{eq:kDowns}
\begin{equation}
\label{eq:kpartDowns}
 \vs{\mu}_k    = \bigsset{\mu\in \Pi(L)}{\abs{\mu}\geq k}.
\end{equation}
(This is related to the natural gradation of the lattice of partitions $\Pi(L)$.)
These form a chain (a completely ordered set),
$\set{\bot}=\vs{\mu}_n
\finereq\dots
\finereq\vs{\mu}_{k+1}
\finereq\vs{\mu}_k
\finereq\dots\finereq\vs{\mu}_1=\Pi(L)$.
($k$-partitionability is related to the $k$-separability in the theory
of multipartite entanglement\cite{Acin-2001,Guhne-2005,Seevinck-2008}:
a mixed state is said to be $k$-separable, if it can be written as the convex mixture of
$k$-partitionably uncorrelated states.
We would call them $k$-partitionably separable
in an extended nomenclature,
describing correlation and entanglement.)
The cuttings corresponding to the nonempty down-sets $\vs{\mu}_k$
are illustrated with dashed green lines in Fig.~\ref{fig:partlattices} 
for $n=2,3,4$ subsystems.

For $k=1,2,\dots,n$, a state is \emph{$k$-producibly uncorrelated},
if it can be written in the product form of density matrices of subsystems of size at most $k$.
That is, they are $\vs{\nu}_k$-uncorrelated for 
the nonempty down-set $\vs{\nu}_k$ containing all the partitions $\nu$
in which the \emph{sizes ($\abs{N}$) of the parts} $N\in\nu$ are less than or equal to $k$,\cite{Szalay-2015b}
\begin{equation}
\label{eq:kprodDowns}
 \vs{\nu}_{k'} = \bigsset{\nu\in \Pi(L)}{\forall N\in\nu: \abs{N}\leq k'}.
\end{equation}
\end{subequations}
(This is related to a ``dual view'' of the lattice of partitions $\Pi(L)$.)
These form a chain, 
$\set{\bot}=\vs{\nu}_1
\finereq\dots
\finereq\vs{\nu}_{k-1}
\finereq\vs{\nu}_k
\finereq\dots\finereq\vs{\nu}_n=\Pi(L)$.
($k$-producibility was originally introduced for the studying of 
multipartite entanglement\cite{Seevinck-2001,Guhne-2005,Toth-2010},
here we use an analogue of that for correlation:
a mixed state is said to be $k$-producible, if it can be written as the convex mixture of
$k$-producibly uncorrelated states.
We would call them $k$-producibly separable
in an extended nomenclature,
describing correlation and entanglement.)
The cuttings corresponding to the nonempty down-sets $\vs{\nu}_k$
are illustrated with dashed orange lines in Fig.~\ref{fig:partlattices} 
for $n=2,3,4$ subsystems.

One can characterise
to what extent a state is not $k$-partitionably and $k$-producibly uncorrelated
by the use of the $\vs{\mu}_k$- and $\vs{\nu}_k$-correlation \eqref{eq:vsxiCorr}.
In this case we call that 
\emph{$k$-partitionability correlation} and
\emph{$k$-producibility correlation}, respectively,
\begin{subequations}
\label{eq:kCorrs}
\begin{align}
\label{eq:kpartCorr}
C_\text{$k$-part}(\varrho_L) &:= C_{\vs{\mu}_k} (\varrho_L) \equiv \min_{\mu \in\vs{\mu}_k }C_\mu (\varrho_L),\\
\label{eq:kprodCorr}
C_\text{$k$-prod}(\varrho_L) &:= C_{\vs{\nu}_k}(\varrho_L) \equiv \min_{\nu\in\vs{\nu}_k}C_\nu(\varrho_L).
\end{align}
\end{subequations}
That is, for an $n$-partite system, we have two groups of $n$ functions measuring correlation. 
Because of the multipartite monotonicity \eqref{eq:MultiMonII},
the $k$-partitionability correlation takes lower value for a smaller $k$,
and
the $k$-producibility correlation takes higher value for a smaller $k$,
\begin{subequations}
\label{eq:MultiMonkpartkprod}
\begin{align}
k\leq k'\quad&\Longleftrightarrow\quad
C_\text{$k$-part}\leq C_\text{$k$'-part},\\
k\leq k'\quad&\Longleftrightarrow\quad
C_\text{$k$-prod}\geq C_\text{$k$'-prod}.
\end{align}
\end{subequations}
Note that $\vs{\mu}_1=\vs{\nu}_n=\Pi(L)$, so
$C_\text{$1$-part}=C_\text{$n$-prod}=0$;
$\vs{\mu}_2=\vs{\nu}_{n-1}$, so 
$C_\text{$2$-part}=C_\text{$(n-1)$-prod}$;
$\vs{\mu}_n=\vs{\nu}_1=\{\bot\}$, so 
$C_\text{$n$-part}=C_\text{$1$-prod}=C_{\{\bot\}}=C_\bot$.
There are no such coincidences
for other values of $k$ in general, however,
$\vs{\mu}_k\finereq \vs{\nu}_{n-k+1}$
(because, for any partition $\xi\in\Pi(L)$,
the relation $\abs{X}\leq n-(\abs{\xi}-1)$ holds for all $X\in\xi$ parts),
so 
\begin{equation}
C_\text{$k$-part}\geq C_\text{$(n-k+1)$-prod},
\end{equation}
because of the multipartite monotonicity \eqref{eq:MultiMonII}.

%%%%%%%%%%%%%%%%%%%%%%%%%%%%%%%%%%%%%%%%%%%%%%%%%%%%%%%%%%%%%%%%%%%%%%%%%%%%%%%%
\subsection{Correlations in subsystems and coarsening.}

Until this point, we considered the correlation measures \eqref{eq:xiCorr}, \eqref{eq:vsxiCorr} and \eqref{eq:kCorrs}
in a system $L=\set{1,2,\dots,n}$ of elementary subsystems
by the use of the partitions 
$\xi =\set{X_1,X_2,\dots,X_{\abs{\xi}} }\equiv X_1|X_2|\dots|X_{\abs{\xi}}\in\Pi(L)$.
There are two ways for making these concepts more flexible.

First,
we would like to characterise the correlations in a (nonempty) subsystem $L'\subseteq L$.
Then for Level I, obviously, we have the $\xi'$-correlation for $\xi'\in\Pi(L')$ with the same definition \eqref{eq:xiCorr}.
For Level II, we also have the $\vs{\xi}'$-correlation for a nonempty down-set $\vs{\xi}'\subseteq\Pi(L')$ with the same definition \eqref{eq:vsxiCorr}.
For the $k$-partitionability and $k$-producibility correlations,
we have to denote the restriction for subsystem,
and we use the notation $C_{\text{$k$-part},L'}(\varrho_{L'})$ and $C_{\text{$k$-prod},L'}(\varrho_{L'})$, respectively.

One can also restrict the notions for (nonempty) subsystems $L'\subseteq L$ from the original system $L$.
Let us denote the restriction of a partition $\xi\in\Pi(L)$
to subsystem $L'$ with $\xi|_{L'}=\sset{X\cap L'\neq\emptyset}{X\in\xi}\in\Pi(L')$.
It is easy to check that 
if $\upsilon\finereq\xi$ then $\upsilon|_{L'}\finereq\xi|_{L'}$.
Let us denote the restriction of a nonempty down-set of partitions $\vs{\xi}$
to subsystem $L'$ with $\vs{\xi}|_{L'}=\sset{\xi|_{L'}}{\xi\in\vs{\xi}}$.
It is easy to check that 
if $\vs{\upsilon}\finereq\vs{\xi}$ then $\vs{\upsilon}|_{L'}\finereq\vs{\xi}|_{L'}$.
The $k$-partitionability and $k$-producibility  of the whole system $L$
is described by the $\vs{\mu}_k$ and $\vs{\nu}_k$ down-sets of partitions in $\Pi(L)$
(see above).
Then, the $k$-partitionability and $k$-producibility of the subsystem $L'$
is described by the $\vs{\mu}_{k,L'}$ and $\vs{\nu}_{k,L'}$ down-sets of partitions in $\Pi(L')$.
We have then that
\begin{subequations}
\label{eq:kpartprodrestrict}
\begin{align}
\label{eq:kpartrestrict}
\vs{\mu}_k|_{L'} &= \vs{\mu}_{k-(\abs{L}-\abs{L'}),L'},\\
\label{eq:kprodrestrict}
\vs{\nu}_k|_{L'} &= \vs{\nu}_{k,L'}.
\end{align}
\end{subequations}
(So, restricting the $k$-partitionability of the system makes sense 
only if $k$ is larger than the number of dropped subsystems plus one;
and restricting the $k$-producibility of the system makes sense 
only if $k$ is smaller than the number of the kept subsystems.)
The proofs of these are as follows.
Notice that $0\leq\abs{\xi}-\abs{\xi|_{L'}}\leq \abs{L\setminus L'}=\abs{L}-\abs{L'}$,
because the number of parts always decreases for restriction,
and the largest decrease occurs 
when all the parts $X\in\xi$ which are not contained in $L'$
are of size $1$, then the number of empty sets, coming from $X\cap L'$, is $\abs{L\setminus L'}$.
Rearranging the relations, we have the bounds on the size of the restricted partition
$\abs{\xi}-\abs{L\setminus L'}\leq\abs{\xi|_{L'}}\leq \abs{\xi}$.
For the proof of \eqref{eq:kpartrestrict}, we need, on the one hand, 
that $\forall \mu\in\vs{\mu}_k$, $\mu|_{L'}\in\vs{\mu}_{k-\abs{L\setminus L'},L'}$.
This comes from that
$k-\abs{L\setminus L'}\leq\abs{\mu}-\abs{L\setminus L'}\leq\abs{\mu|_{L'}}$,
where the first inequality is $\abs{\mu}\geq k$ (by \eqref{eq:kpartDowns} definition of $\vs{\mu}_k$)
and the second one is the above bound on the size of the restricted partition.
For the proof of \eqref{eq:kpartrestrict}, we need, on the other hand, 
that $\forall\mu'\in \vs{\mu}_{k-\abs{L\setminus L'},L'}$, $\exists\mu\in\vs{\mu}_k$
such that $\mu'=\mu|_{L'}$.
For this,
we consider the partition $\mu=\mu'\cup\mu''\in\Pi(L)$,
where $\mu''\in\Pi(L\setminus L')$.
Then $\mu|_{L'}=\mu'$ holds clearly.
Let $\mu''$ be the bottom element of $\Pi(L\setminus L')$,
that is, $\mu'':=\bot_{\Pi(L\setminus L')}=\sset{\{i\}}{i\in L\setminus L'}$,
then $\abs{\mu}=\abs{\mu'}+\abs{\bot_{\Pi(L\setminus L')}}=\abs{\mu'}+\abs{L\setminus L'}\geq k$,
so $\mu\in\vs{\mu}_k$.
For the proof of \eqref{eq:kprodrestrict}, we need, on the one hand,
$\forall \nu\in\vs{\nu}_k$, $\nu|_{L'}\in\vs{\nu}_{k,L'}$.
This comes from the \eqref{eq:kprodDowns} definition of $\vs{\nu}_k$,
and from that the size of the parts of $\nu$ decreases for restriction.
For the proof of \eqref{eq:kprodrestrict}, we need, on the other hand,
that $\forall\nu'\in \vs{\nu}_{k,L'}$, $\exists\nu\in\vs{\nu}_k$
such that $\nu'=\nu|_{L'}$.
For this,
we consider the partition $\nu=\nu'\cup\nu''\in\Pi(L)$,
where $\nu''\in\Pi(L\setminus L')$.
Then $\nu|_{L'}=\nu'$ holds clearly.
Let $\nu''$ be also $k$-producible, that is, $\nu''\in\vs{\nu}_{k,L\setminus L'}$
then clearly $\nu\in\vs{\nu}_k$, by \eqref{eq:kprodDowns} definition of $\vs{\nu}_k$.

Second,
we would like to characterise the correlations in the whole system $L$, but only with respect to a coarsening,
given by the partition $\upsilon=Y_1|Y_2|\dots|Y_{\abs{\upsilon}}\in\Pi(L)$.
That is, the composite subsystems $Y\in\upsilon$ become the elementary ones in the coarsened treatment.
This means that, for the correlations, we use the partitions coarser than $\upsilon$, 
the set of which is denoted by 
$\Pi(L,\upsilon)=\sset{\xi'\in\Pi(L)}{\upsilon\finereq\xi'}=\upset\{\upsilon\}\subseteq\Pi(L)$,
which is a principal filter\cite{Davey-2002} in $\Pi(L)$.
Then for Level I, we have the $\xi'$-correlation for $\xi'\in\Pi(L,\upsilon)$ with the same definition \eqref{eq:xiCorr}.
For Level II, we also have the $\vs{\xi}'$-correlation for a nonempty down-set $\vs{\xi}'\subseteq\Pi(L,\upsilon)$ with the same definition \eqref{eq:vsxiCorr}.
For the $k$-partitionability and $k$-producibility correlations,
we have to denote the coarsening
and we use the notation $C_{\text{$k$-part},\upsilon}(\varrho_L)$ and $C_{\text{$k$-prod},\upsilon}(\varrho_L)$, respectively.
(Note that in the case of $k$-producibility, the minimisation in \eqref{eq:kprodCorr} should be taken
over the partitions $\nu\in\vs{\nu}_{k,\upsilon}=\vs{\nu}_k\cap\Pi(L,\upsilon)  \subseteq\Pi(L,\upsilon)$
in which the parts $N\in\nu$ are the disjoint union of subsystems $Y\in\upsilon$ of number less than or equal to $k$.)

%%%%%%%%%%%%%%%%%%%%%%%%%%%%%%%%%%%%%%%%%%%%%%%%%%%%%%%%%%%%%%%%%%%%%%%%%%%%%%%%
\subsection{Global bounds.}

The von Neumann entropy \eqref{eq:vonNeumann} takes its maximum for maximally mixed states,\cite{Ohya-1993}
so it is bounded by
\begin{equation}
\label{eq:vonNeumannBound}
S(\varrho_X)\leq\ln d_X = \abs{X}\ln d,
\end{equation}
where $d$ is the (uniform) dimension of the Hilbert spaces of the elementary subsystems.

For disjoint $X,X'\in L$ subsystems, 
we have that the usual bipartite $X|X'$-correlation \eqref{eq:xiCorr}
is bounded by
\begin{equation}
C_{X|X'}(\varrho_{X\cup X'})\leq\bigl(\abs{X}+\abs{X'}\bigr)\ln d.
\end{equation}
Another bound can also be given,
which is stronger, if one part is larger than the other,
\begin{equation}
\label{eq:MutInfBound}
\begin{split}
C_{X|X'}(\varrho_{X\cup X'})
&\leq \min\bigset{2\ln d_X, 2\ln d_{X'}}\\
&= \min\bigset{\abs{X},\abs{X'}}2\ln d.
\end{split}
\end{equation}
The proof of this is as follows,
\begin{equation*}
\begin{split}
C_{X|X'}(\varrho_{X\cup X'})
&=    S(\varrho_X) + S(\varrho_{X'}) - S(\varrho_{X\cup X'})\\
&\leq S(\varrho_X) + S(\varrho_{X'}) - \bigabs{S(\varrho_X) - S(\varrho_{X'})}\\
&=    2\min\bigset{S(\varrho_X),S(\varrho_{X'})},
\end{split}
\end{equation*}
where the \eqref{eq:vonNeumannTriangle} Araki-Lieb triangle inequality,
is used.
Then using \eqref{eq:vonNeumannBound} completes the proof:
the minimum of the entropies is bounded from above by the minimum of the upper bounds of the entropies.

Generalising the above,
let us consider a partition $\xi=X_1|X_2|\dots|X_{\abs{\xi}}\in\Pi(L)$ of the whole system $L$.
We have then that the $\xi$-correlation \eqref{eq:xiCorr}
is bounded by
\begin{equation}
C_\xi(\varrho_L)\leq \abs{L}\ln d.
\end{equation}
Another bound can also be given,
which is stronger, if one part is larger than the others together,
\begin{equation}
\label{eq:xiCorrBound}
\begin{split}
C_\xi(\varrho_L)
&\leq  \Bigl(\sum_{X\in\xi}2\ln d_X - \max_{X'\in\xi}\bigset{ 2\ln d_{X'}}  \Bigr)\\
&= \Bigl( \abs{L} - \max_{X\in\xi} \bigset{ \abs{X} } \Bigr)2\ln d.
\end{split}
\end{equation}
The proof of this is as follows,
\begin{equation*}
\begin{split}
C_\xi(\varrho_L) 
&=    \sum_{X\in\xi}S(\varrho_X) - S(\varrho_L)\\
&\leq \sum_{X\in\xi}S(\varrho_X) -  S(\varrho_{X'}) + \sum_{\substack{X\in\xi\\X\neq X'}}S(\varrho_X)\\
&= 2\sum_{\substack{X\in\xi\\X\neq X'}}S(\varrho_X)\qquad\text{for all $X'\in\xi$},
\end{split}
\end{equation*}
where the polygon inequality,
\begin{equation*}
S(\varrho_{X'}) - \sum_{\substack{X\in\xi\\X\neq X'}}S(\varrho_X)\leq S(\varrho_L) \qquad\text{for all $X'\in\xi$},
\end{equation*}
is used.
(This follows from the combination of the \eqref{eq:vonNeumannTriangle} triangle inequality
in the form of $S(\varrho_{X'}) - S(\varrho_{L\setminus X'}) \leq S(\varrho_L)$,
holds for all $X'\subseteq L$,
with the \eqref{eq:vonNeumannSA} subadditivity,
in the form $S(\varrho_{L\setminus X'})\leq\sum_{X\in\xi,X\neq X'}S(\varrho_X)$.)
Then using \eqref{eq:vonNeumannBound} completes the proof:
the sum of the $\abs{\xi}-1$ lowest entropies is bounded from above by the sum of the $\abs{\xi}-1$ lowest upper-bound of the entropies.
Note that \eqref{eq:MutInfBound} is a special case of this.

Having the \eqref{eq:xiCorrBound} bound for the $\xi$-correlation \eqref{eq:xiCorr},
we can obtain the bound 
\begin{equation}
\label{eq:vsxiCorrBound}
C_{\vs{\xi}}(\varrho_L) \leq \Bigl( \abs{L} - \max_{\xi\in\vs{\xi}} \max_{X\in\xi} \bigset{ \abs{X} } \Bigr)2\ln d
\end{equation}
for the $\vs{\xi}$-correlation \eqref{eq:vsxiCorr}:
the minimum of $\xi$-correlations is bounded from above by the minimum of the upper-bounds of the $\xi$-correlations.
This leads to the bounds for 
the $k$-partitionability correlation \eqref{eq:kpartCorr} and
the $k$-producibility correlation \eqref{eq:kprodCorr}
\begin{subequations}
\label{eq:kpartprodMutInfBound}
\begin{align}
\label{eq:kpartMutInfBound}
C_\text{$k$-part}(\varrho_L) &\leq \bigl(k-1\bigr)2\ln d,\\
\label{eq:kprodMutInfBound}
C_\text{$k$-prod}(\varrho_L) &\leq \bigl(\abs{L}-k\bigr)2\ln d.
\end{align}
\end{subequations}

Based on the bounds \eqref{eq:MutInfBound}, \eqref{eq:xiCorrBound}, \eqref{eq:vsxiCorrBound} and \eqref{eq:kpartprodMutInfBound},
it is convenient to give all numerical results for these quantities in units of $\ln d$.
On the other hand, although we do not know 
if the bounds \eqref{eq:xiCorrBound}, \eqref{eq:vsxiCorrBound} and \eqref{eq:kpartprodMutInfBound} 
can be attained or not by quantum states,
we emphasise that 
there exist stronger bounds for the correlations in classical states 
than for the correlations in quantum states.
Indeed, applying the monotonicity of the entropy w.r.t.~partial trace
($S(\varrho_{X'})\leq S(\varrho_X)$ if $X'\subseteq X$,
holds for classical states\cite{Ohya-1993}),
it is easy to prove, that for classical states
we have half of the bounds \eqref{eq:MutInfBound}, \eqref{eq:xiCorrBound}, \eqref{eq:vsxiCorrBound} and \eqref{eq:kpartprodMutInfBound}.
For example, consider the cases of two classical and two quantum dits.
For classical states (embedded \emph{locally} into quantum states), the maximally correlated one 
is given by $\frac1d\sum_{i=1}^d\cket{i}\bra{i}\otimes\cket{i}\bra{i}$, for which $C_{1|2}=\ln d$,
while for quantum states, the maximally correlated one 
is given by $\frac1d\sum_{i,j=1}^d\cket{i}\bra{j}\otimes\cket{i}\bra{j}$
(projecting to the state vector $\frac1{\sqrt{d}}(\sum_{i=1}^d\cket{i}\otimes\cket{i}$), for which $C_{1|2}=2\ln d$.
That is, the \eqref{eq:MutInfBound} bound is strict for quantum states (i.e., it is attained),
and the (strict) bound for classical states is only the half of that.
Similarly, we expect that the maximal possible values of the  $\xi$-correlation, the $\vs{\xi}$-correlation, 
the $k$-partitionability correlation and the $k$-producibility correlation are strictly smaller for classical states than for quantum states.

%%%%%%%%%%%%%%%%%%%%%%%%%%%%%%%%%%%%%%%%%%%%%%%%%%%%%%%%%%%%%%%%%%%%%%%%%%%%%%%%
\subsection{Relations among the correlation measures.}

We have already seen the first two relations among the correlation measures,
the multipartite monotonicity of the first and second kinds in \eqref{eq:MultiMonI} and \eqref{eq:MultiMonII}.
We can easily have the further two,
\begin{equation}
\label{eq:vsxixi}
C_{\vs{\xi}}(\varrho_L)\leq C_\xi(\varrho_L)\qquad\text{for all $\xi\in\vs{\xi}$},
\end{equation}
by definition \eqref{eq:vsxiCorr},
and 
\begin{equation}
\label{eq:xivsxi}
C_\xi(\varrho_L)\leq C_{\vs{\xi}}(\varrho_L)\qquad\text{if $\vs{\xi}\finereq\downset\{\xi\}$},
\end{equation}
by the multipartite monotonicity \eqref{eq:MultiMonII}.

For disjoint $X,X'\in L$ subsystems,
we have that the usual bipartite $X|X'$-correlation
is monotonic in the sense that
\begin{equation}
\label{eq:edgeboundbipart}
\forall i\in X,i'\in X':\quad C_{i|i'}(\varrho_{\set{i,i'}})\leq C_{X|X'}(\varrho_{X\cup X'}), 
\end{equation}
following from \eqref{eq:geomCorrImin} 
and the \eqref{eq:UmegakiRedMon} monotonicity of the relative entropy \eqref{eq:Umegaki}.
(Alternatively, a different proof can be formulated exploiting the \eqref{eq:vonNeumannSSA} strong subadditivity of the von Neumann entropy.)
From this, and the multipartite monotonicity \eqref{eq:MultiMonI}, for all $i,i'$ orbitals and for all $\xi\in\Pi(L)$ splits,
\begin{equation}
\label{eq:edgeboundI}
C_{i|i'}(\varrho_{\set{i,i'}})
\leq C_\xi(\varrho_L)\qquad
\text{if $i|i'=\xi|_{\{i,i'\}}$,}
\end{equation}
that is, when $\xi$ separates $i$ and $i'$.
From this, 
\begin{equation}
\label{eq:edgeboundII}
\begin{split}
&C_{i|i'}(\varrho_{\set{i,i'}})
\leq \min_{\xi\in\vs{\xi}}C_\xi(\varrho_L)
=C_{\vs{\xi}}(\varrho_L)\\
&\qquad\text{if $i|i'=\xi_*|_{\{i,i'\}}$, where $\xi_*=\argmin_{\xi\in\vs{\xi}}C_\xi(\varrho_L)$.} %\quad 
\end{split}
\end{equation}
In particular, 
\begin{equation}
\label{eq:edgebound2part}
\begin{split}
&C_{i|i'}(\varrho_{\set{i,i'}})
\leq \min_{X\subset L}C_{X|(L\setminus X)}(\varrho_L)
=C_\text{$2$-part}(\varrho_L)\\
&\qquad\text{if $i\in X_*,i'\in L\setminus X_*$,}\\
&\qquad\text{where $X_*=\argmin_{X\subset L}C_{X|(L\setminus X)}(\varrho_L)$.} %\quad 
\end{split}
\end{equation}

Generalising the relations above,
let us consider a (nonempty) subsystem $L'\subseteq L$ of the whole system $L$,
and the $\xi|_{L'}$ restriction of a partition $\xi$ to this subsystem.
We have then
\begin{equation}
\label{eq:restrictboundI}
C_{\xi|_{L'}}(\varrho_{L'})\leq C_\xi(\varrho_L),
\end{equation}
following from \eqref{eq:geomCorrImin} and the monotonicity of the relative entropy \eqref{eq:Umegaki} for partial trace\cite{Petz-2008,Wilde-2013}.
Note that the bound \eqref{eq:edgeboundI} is a special case of this.
Let us consider the $\vs{\xi}|_{L'}$ restriction of a nonempty down-set of partitions $\vs{\xi}$ to this subsystem.
We have then 
\begin{equation}
\label{eq:restrictboundII}
C_{\vs{\xi}|_{L'}}(\varrho_{L'})
\leq C_{\vs{\xi}}(\varrho_L),
\end{equation}
following from \eqref{eq:restrictboundI}:
the minimum of the $\xi|_{L'}$-correlations is bounded from above by the
minimum of the larger $\xi$-correlations.
Let us consider the $k$-partitionability and $k$-producibility of the subsystem $L'$.
We have then
\begin{subequations}
\begin{align}
C_{\text{$\bigl(k-(\abs{L}-\abs{L'})\bigr)$-part},L'}(\varrho_{L'}) &\leq C_\text{$k$-part}(\varrho_L),\\
C_{\text{$k$-prod},L'}(\varrho_{L'}) &\leq C_\text{$k$-prod}(\varrho_L),
\end{align}
\end{subequations}
because of \eqref{eq:kpartprodrestrict}.

On the other hand,
one can also bound level II measures of subsystems $L'\subseteq L$ by level I measures of the original system $L$.
We have then that for a $\xi\in\Pi(L)$ split,
\begin{equation}
\label{eq:combinedboundII}
C_{\vs{\xi}'}(\varrho_{L'})
\leq C_\xi(\varrho_L)\qquad
\text{if $\xi|_{L'}\in\vs{\xi}'$},
\end{equation}
following from \eqref{eq:restrictboundI} and \eqref{eq:vsxixi}.
In particular,
\begin{equation}
\label{eq:combinedbound2part}
\begin{split}
C_{\text{$2$-part},L'}(\varrho_{L'})
&\leq C_\xi(\varrho_L)\\
&\text{if $\xi|_{L'}\finereq X'|(L'\setminus X')$ for a $X'\in L'$,}
\end{split}
\end{equation}
that is, if $\xi$ dissects $L'$,
that is, if $\xi|_{L'}$ is not the trivial split.

%%%%%%%%%%%%%%%%%%%%%%%%%%%%%%%%%%%%%%%%%%%%%%%%%%%%%%%%%%%%%%%%%%%%%%%%%%%%%%%%
\subsection{Relations for the bipartite correlation clustering.}

Let us split the system into subsystems, 
described by the partition $\gamma=G_1|G_2|\dots|G_{\abs{\gamma}}\in\Pi(L)$,
given by the clustering based on the ``connectivity''
with respect to $C_{i|j}$.
That is, subsystems $i$ and $j$ are contained in the same part $G\in\gamma$,
if and only if there exists a path $i=i_1,i_2,\dots,i_p=j$ of orbitals
for which $C_{i_s|i_{s+1}}(\varrho_{\{i_s,i_{s+1}\}})\geq T_\text{b}$ for a threshold $T_\text{b}$ for all $1\leq s \leq p-1$.
We call this \emph{bipartite correlation clustering}.

The first point to see here is that
there are nonvanishing 
correlations $C_{\xi'}$ and $C_{\vs{\xi}'}$ inside the parts $G\in\gamma$.
Indeed, for all $\xi'\in\Pi(G)$, for all $X'\in\xi'$,
there are $i'\in X'$ and $i\in G\setminus X'$ for which
$C_{i|i'}(\varrho_{\{i,i'\}})\geq T_\text{b}$, because of the construction of $\gamma$.
For these, on the other hand, $\xi'|_{\{i,i'\}}=i|i'$,
so the condition of \eqref{eq:edgeboundI} holds,
\begin{equation}
\label{eq:edgeboundbondcorrI}
T_\text{b} \leq C_{i|i'}(\varrho_{\{i,i'\}})\leq C_{\xi'}(\varrho_G).
\end{equation}
Because this holds for all $\xi'$, 
it holds also for $\xi'_*=\argmin_{\xi'\in\vs{\xi}'}C_{\xi'}$,
so the condition of \eqref{eq:edgeboundII} holds,
\begin{equation}
\label{eq:edgeboundbondcorrII}
T_\text{b}\leq C_{i|i'}(\varrho_{\{i,i'\}})\leq C_{\vs{\xi}'}(\varrho_G).
\end{equation}
In particular,
\begin{equation}
\label{eq:edgeboundbondcorr2part}
T_\text{b}\leq C_{i|i'}(\varrho_{\{i,i'\}})\leq C_{\text{$2$-part},G}(\varrho_G).
\end{equation}
Note that, because of \eqref{eq:xiCorrBound} and \eqref{eq:vsxiCorrBound},
the bounds \eqref{eq:edgeboundbondcorrI} and \eqref{eq:edgeboundbondcorrII}
seem to be rather weak.
However, because of \eqref{eq:kpartMutInfBound},
the bound \eqref{eq:edgeboundbondcorr2part} seems to be strong, depending on $T_\text{b}$.

The second point to see here is that
$C_\gamma$ is not necessarily weak.
If there is a subsystem $L'\subseteq L$ which is dissected by $\gamma$,
that is, for which $\gamma|_{L'}$ is nontrivial,
and the correlation $C_{\vs{\xi}'}$ inside $L'$
is strong
(we can interpret this as the occurrence of \emph{hidden correlations}),
then $C_\gamma$ is also strong,
\begin{equation}
\label{eq:bondboundII}
C_{\vs{\xi}'}(\varrho_{L'})\leq C_\gamma(\varrho_L)\qquad\text{if $\gamma|_{L'}\in\vs{\xi}'$},
\end{equation}
which is \eqref{eq:combinedboundII}.
In particular,
\begin{equation}
\label{eq:bondbound2part}
C_{\text{$2$-part},L'}(\varrho_{L'})\leq C_\gamma(\varrho_L)
\end{equation}
in all cases, because all nontrivial splits are contained in $\vs{\mu}_{2,L'}$.

\begin{figure}
\centering
% \fbox{
\includegraphics{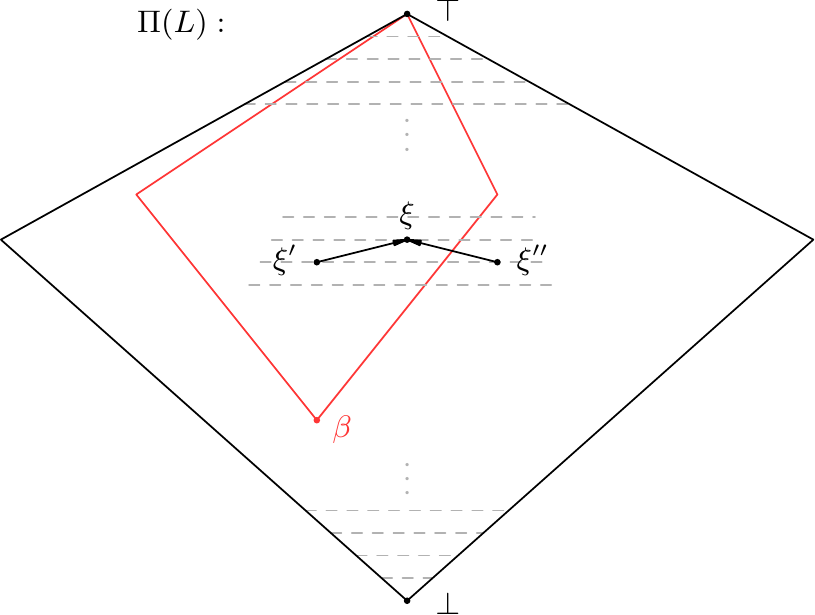}
% } % of \fbox
\caption{Illustration for the \eqref{eq:mcclocal} definition of multipartite correlation clustering.
The up-set $\upset\{\beta\}$ is drawn by red line,
$C_{\xi'}-C_\xi\leq T_\text{m}$, because $\beta\preceq\xi'$, while
$C_{\xi''}-C_\xi> T_\text{m}$, because $\beta\npreceq\xi''$.
(Schematic view of the partition lattice $\Pi(L)$:
the arrows denote the covering relation \eqref{eq:coveringParts} in the same way as in Fig.~\ref{fig:partlattices},
and the dashed grey lines represent the gradation of the lattice.)}
\label{fig:cldef}
\end{figure}

The third point to see here is that 
if $C_\xi$ is weak for a split $\xi\in\Pi(L)$, then $\gamma\finereq\xi$,
or, contrapositively,
if $\gamma\nfinereq\xi$ then $C_\xi$ is strong.
This comes as follows.
If $\gamma\nfinereq\xi$ then there is a $G\in\gamma$ which is dissected by $\xi$,
that is, $\xi':=\xi|_G$ is not trivial.
Then for all $X'\in\xi'$, 
there are $i'\in X'$ and $i\in G\setminus X'$ for which
$C_{i|i'}(\varrho_{\{i,i'\}})\geq T_\text{b}$, because of the construction of $\gamma$.
For these, on the other hand, $\xi'|_{\{i,i'\}}=i|i'$,
so the condition of \eqref{eq:edgeboundI} holds,
and using also \eqref{eq:restrictboundI}, we have
$T_\text{b}\leq C_{i|i'}(\varrho_{\{i,i'\}}) \leq C_{\xi'}(\varrho_G)\leq C_\xi(\varrho_L)$.

The fourth point to see here is that
one can exclude the hidden correlations among the parts of $\xi\in\Pi(L)$ if $C_\xi$ is weak.
We have just seen that if $C_\xi$ is weak then $\gamma\finereq\xi$.
If $C_\xi$ is weak then the $2$-partitionability $C_{\text{$2$-part},L'}(\varrho_{L'})$
is weak in every subsystem $L'\subset L$ which is dissected by $\xi$,
that is, for which $\xi|_{L'}$ is nontrivial, that is,
\begin{equation}
\label{eq:hiddenexclude}
C_{\text{$2$-part},L'}(\varrho_{L'})\leq C_\xi(\varrho_L),
\end{equation}
in the same way as \eqref{eq:bondbound2part},
and here the right hand side is weak.
However, if the system is large enough,
then there can be several small local contributions to the global $C_\xi$, 
making it too large, even if there are no hidden correlations.

In summary, an intrinsic problem of the bipartite correlation clustering is that 
it is based on bipartite correlations,
which are local (that is, consider only density matrices of two elementary subsystems), 
and which ranges in $0\leq C_{i|j}\leq 2\ln d$.
Because of this, it is unable to grasp the multipartite correlations in a satisfactory way,
unless some additivity results for $C_{i|j}$ can be proven, which does not seem to be the case.

\begin{figure*}
\centering
\setlength{\unitlength}{12.6bp}   % u = 13bp in the metapost file
 \setlength{\fboxsep}{0pt}
% \fbox{
\begin{picture}(40,14.5)

\put(00.0,00.0){\includegraphics{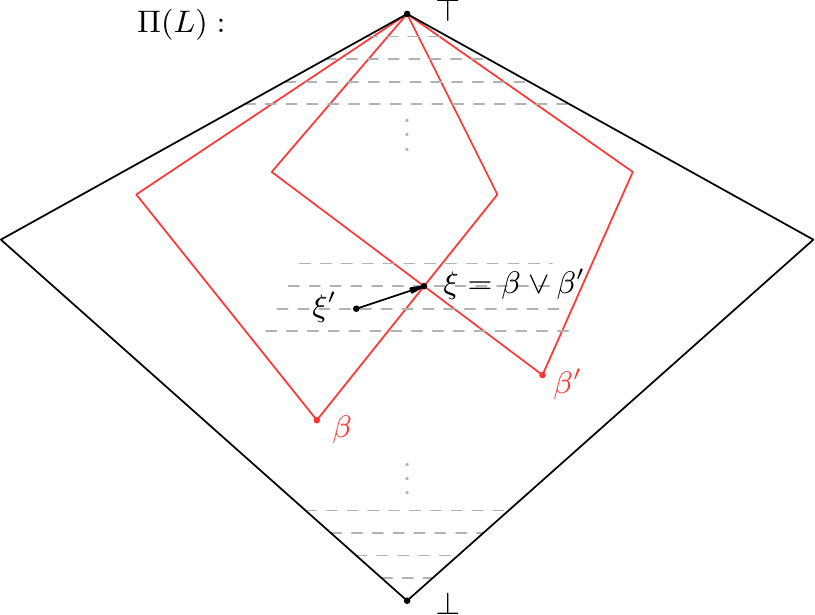}}
\put(20.0,00.0){\includegraphics{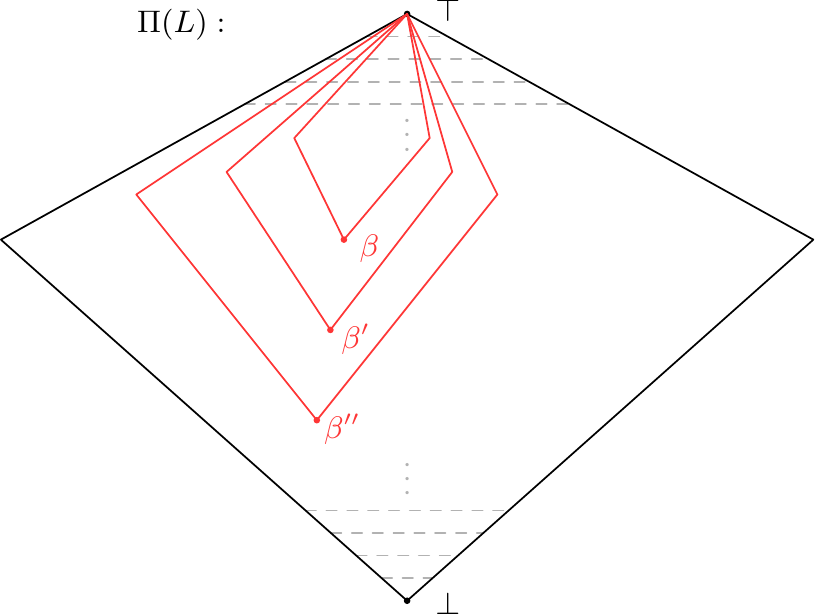}}

\put(00.0, 14.0){\makebox(0,0)[r]{\strut{}}(a)}
\put(20.0, 14.0){\makebox(0,0)[r]{\strut{}}(b)}

\end{picture}
% } % of \fbox
\caption{Illustrations for the multipartite correlation clusterings.
(a) There are no contradictory multipartite correlation clusterings.
(b) There may be different compatible multipartite correlation clusterings.}
\label{fig:clcc}
\end{figure*}

%%%%%%%%%%%%%%%%%%%%%%%%%%%%%%%%%%%%%%%%%%%%%%%%%%%%%%%%%%%%%%%%%%%%%%%%%%%%%%%e
\subsection{Multipartite correlation clustering}

Here we formulate and solve the task
of dividing the whole system
into weakly correlated subsystems consisting of strongly correlated elementary subsystems.
We call this \emph{multipartite correlation clustering}.
That is, to obtain $\beta=B_1|B_2|\dots|B_{\abs{\beta}}\in\Pi(L)$, \emph{if exists,} for which\\[5pt]
(i) the subsystems described by the parts $B\in\beta$ are weakly correlated with one another,\\
(ii) the elementary subsystems $i\in B$ inside a part $B\in\beta$ are strongly correlated with one another.\\[5pt]
It makes this notion complicated that
strong and weak are ill-defined, and depend on the context.
That is, although some rules of thumb might exist,
 we cannot formulate general thresholds for
$C_{\text{$k$-part},B}$, $C_{\text{$k$-prod},B}$ and $C_\beta$ independently of the situation.
Instead of that, we use a different point of view, leading to a \emph{local} strategy.
For this,
we have to be able to decide
about a given $\xi$, 
if it is a good ansatz, or it is worth considering a $\xi'$,
which is ``a bit'' finer than $\xi$.

The first we need is
to calculate the ``derivative'' of $C_\xi$ with respect to $\xi$,
that is, the difference of $C_\xi$ for covering $\xi$ values.
Let $\xi'\covered\xi$,
then we have
\begin{equation}
\label{eq:deltaC}
C_{\xi'}(\varrho_L)-C_\xi(\varrho_L) = C_{\xi'\setminus\xi}(\varrho_X),
\end{equation}
where $X$ is the unique element in $\xi\setminus\xi'$, see \eqref{eq:coveringParts}.
Indeed, if $\xi'\covered\xi$, then
they have the same parts, the entropies of which cancel each other, 
apart that there is a unique $X_*\in\xi$, which is dissected into
the disjoint $X_{*1}',X_{*2}'\in\xi'$ parts, from which 
$C_{\xi'}(\varrho_L)-C_\xi(\varrho_L) =S(\varrho_{X_{*1}'})+S(\varrho_{X_{*2}'})-S(\varrho_{X_*})
=C_{X_{*1}'|X_{*2}',X_*}(\varrho_{X_*})$.
(Note that the right hand side in \eqref{eq:deltaC} is nonnegative, 
so $C_\xi$ decreases with respect to the covering,
which is the special case of the multipartite monotonicity \eqref{eq:MultiMonI},
valid for arbitrary coarsening.)

Now we reformulate the multipartite correlation clustering (i)-(ii)
as seeking $\beta$ for which
\begin{equation}
\label{eq:mcclocal}
\begin{split}
&\text{there exists a threshold $T_\text{m}>0$, such that}\\
&\text{$\forall \xi,\xi'\in\Pi(L)$ such that
$\xi'\covered\xi$,
and $\beta\finereq\xi$, then}\\
&\qquad\beta\finereq \xi'\quad\Leftrightarrow\quad C_{\xi'}(\varrho_L)-C_\xi(\varrho_L) \leq T_\text{m}.
\end{split}
\end{equation}
(For illustration, see Fig.~\ref{fig:cldef}.)
This means that, 
on the one hand, the change of the function $C_\xi$ with $\xi$ is small
while $\xi$ is coarser than $\beta$
($\xi$ does not leave the up-set $\upset\{\beta\}$),
that is, we divide only parts weakly correlated with one another \eqref{eq:deltaC}.
This is how (i) is grasped.
On the other hand,
the function $C_\xi$ jumps
when $\xi$ gets not coarser than $\beta$
($\xi$ does leave the up-set $\upset\{\beta\}$),
that is, if we divide parts strongly correlated with one another \eqref{eq:deltaC}.
This is how (ii) is grasped.
Note that, for a more robust definition, one can impose a threshold interval instead of a simple threshold value.
Note also that the minimal change in $C_\xi$ 
is related to the $2$-partitionability of the parts in $\xi$, 
\begin{equation}
\label{eq:stepbound}
\min_{\xi'\covered\xi}\bigl(C_{\xi'}(\varrho_L)-C_\xi(\varrho_L)\bigr) = \min_{X\in\xi}C_{\text{$2$-part},X}(\varrho_X),
\end{equation}
because of \eqref{eq:deltaC}.

\begin{figure*}
\centering
\setlength{\unitlength}{12.6bp}   % u = 13bp in the metapost file
 \setlength{\fboxsep}{0pt}
% \fbox{
\begin{picture}(40,14.5)

\put(00.0,00.0){\includegraphics{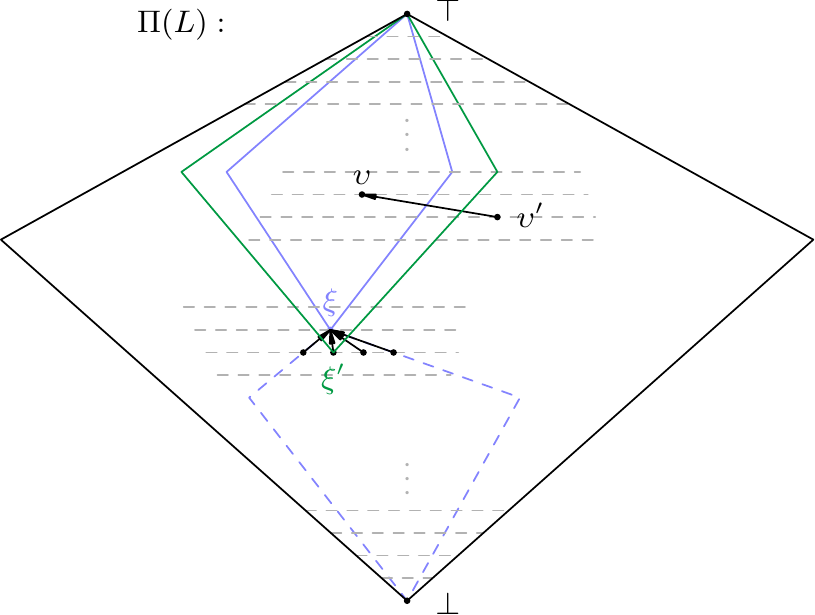}}
\put(25.0,01.0){\includegraphics{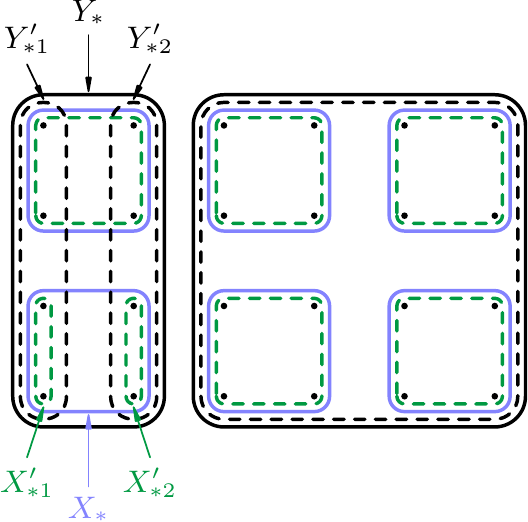}}

\put(00.0, 14.0){\makebox(0,0)[r]{\strut{}}(a)}
\put(20.0, 14.0){\makebox(0,0)[r]{\strut{}}(b)}

\end{picture}
% } % of \fbox
\caption{Illustration for the proof of \eqref{eq:valley} for multipartite correlation clustering.
(a) The up-set $\upset\{\xi\}$ and down-set $\downset\{\xi\}$ are drawn by solid and dashed blue lines.
$\upsilon'$ leaves $\upset\{\xi\}$;
$\xi'$, constructed in the text, is covered by $\xi$,
its up-set, drawn by solid green line, may or may not contain $\upsilon'$.
(b) The construction of $\xi'$, see in the text:
$\upsilon$ and $\upsilon'$ are drawn by solid and dashed black lines.
$\xi$ and $\xi'$ are drawn by solid blue and dashed green lines.}
\label{fig:cltrick}
\end{figure*}

There might not exist meaningful multipartite correlation clustering 
for a given quantum state $\varrho_L$,
that is, there might exist no $\beta$ satisfying \eqref{eq:mcclocal}.
However, the notion of the multipartite correlation clustering
(accordingly, the definition \eqref{eq:mcclocal})
is strong enough to
exclude the existence of more than one \emph{contradictory} $\beta$s
(that is, which are not related by coarsening).
Indeed, let us take, contrapositively,
that we have $\beta$ and $\beta'$, with possibly different thresholds $T_\text{m}\leq T_\text{m}'$,
for which $\beta\npreceq\beta'$ and $\beta\nsucceq\beta'$.
(For illustration, see Fig.~\ref{fig:clcc}(a).)
Now let us have $\xi=\beta\vee\beta'$, the least upper bound of $\beta$ and $\beta'$,
then for the next step $\xi'\covered\xi$ we have that either $\beta\nprec\xi'$ or $\beta'\nprec\xi'$
(it leaves either $\upset\{\beta\}$ or $\upset\{\beta'\}$).
Let us choose a step $\beta'\nprec\xi'$ and $\beta\prec\xi'$ 
(leaving $\upset\{\beta'\}$ but staying in $\upset\{\beta\}$),
then from the definition \eqref{eq:mcclocal} we have that
$T_\text{m}'< C_{\xi'}(\varrho_L)-C_\xi(\varrho_L) \leq T_\text{m}$,
contradicting to $T_\text{m}\leq T_\text{m}'$.
On the other hand, 
more than one \emph{compatible} $\beta$ (with different thresholds $T_\text{m}$) might still exist
(that is, which are related by coarsening).
Then they form a chain, $\beta\coarser\beta'\coarser\beta''\coarser\dots$
($\upset\{\beta\}\subset\upset\{\beta'\}\subset\upset\{\beta''\}\subset\dots$),
and $T_\text{m}< T_\text{m}'< T_\text{m}''< \dots$,
this means that there are different meaningful levels 
of the multipartite correlation clustering,
that is, different strength-scales of correlations.
(For illustration, see Fig.~\ref{fig:clcc}(b).)

How to find $\beta$ satisfying \eqref{eq:mcclocal}?
The threshold $T_\text{m}$ in \eqref{eq:mcclocal} seems also to be ill-defined,
unless we calculate and compare $C_\xi$ for all $\xi\in\Pi(L)$, which
is infeasible even for not too large systems.
($\abs{\Pi(L)}$ grows rapidly\cite{oeisA000110} with $\abs{L}$.)
Fortunately, we do not have to do so.
It is enough to start with the trivial element $\xi=\top=\{L\}$
(with $C_\top=0$),
then carrying out \emph{successive refinement} 
(that is, climbing down $\Pi(L)$ such that
in each step we move from $\xi$ to $\xi'$ which is covered by $\xi$),
while keeping track of the change of $C_\xi$.
If in one step $C_{\xi'}-C_\xi$ is much larger than before, then
we have to find another $\xi'\covered\xi$,
for which $C_{\xi'}-C_\xi$ is small.
If there is no such $\xi'$, then we have reached $\beta$, that is, $\beta=\xi$.
This is because
in general,
\begin{equation}
\label{eq:valley}
\begin{split}
&\text{$\forall \upsilon,\upsilon'\in\Pi(L)$ such that $\upsilon'\covered\upsilon$, and}\\
&\text{$\forall \xi\in\Pi(L)$ such that $\xi\preceq\upsilon$ but $\xi\npreceq\upsilon'$, then}\\
&\qquad\min_{\xi'\covered\xi} C_{\xi'}(\varrho_L)-C_\xi(\varrho_L) \leq
C_{\upsilon'}(\varrho_L)-C_\upsilon(\varrho_L).
\end{split}
\end{equation}
(For illustration, see Fig.~\ref{fig:cltrick}(a).)
The proof is as follows.
For the right hand side,
since $\upsilon'\covered\upsilon$,
we have the unique $Y_*\in\upsilon$ and $Y_{*1}',Y_{*2}'\in\upsilon'$, such that
$\upsilon'\setminus\upsilon = \{Y_{*1}',Y_{*2}'\}$  and
$\upsilon\setminus\upsilon' = \{Y_*\}$, see \eqref{eq:coveringParts},
by which 
$C_{\upsilon'}(\varrho_L)-C_\upsilon(\varrho_L) = C_{\{Y_{*1}',Y_{*2}'\},Y_*}(\varrho_{Y_*})$, see \eqref{eq:deltaC}.
For the left hand side,
since $\xi\preceq\upsilon$,
we have that for all $X\in\xi$ there exists a $Y\in\upsilon$ such that $X\subseteq Y$, see \eqref{eq:oderingParts}.
The partition $\upsilon'$ dissects some (at least one) parts of $\xi$
(since $\xi\npreceq\upsilon'$).
Let $X_*\in\xi$ be such a part.
Note that $X_*\subseteq Y_*$, since $\xi\preceq\upsilon$, and $Y_*$ is the one dissected by $\upsilon'$.
Now choose a $\xi'$ such that 
$X_*\in\xi$ is also dissected by $\xi'$ into parts ``in the same way as $\upsilon'$'',
that is,
$\xi':=\bigl(\xi\setminus\{X_*\}\bigr) \cup \{X_{*1}',X_{*2}'\}$
where
$X_{*1}':=X_*\cap Y_{*1}'\subseteq Y_{*1}'$ and 
$X_{*2}':=X_*\cap Y_{*2}'\subseteq Y_{*2}'$.
(For illustration, see Fig.~\ref{fig:cltrick}(b).)
It is clear that $\xi'\covered\xi$, see \eqref{eq:coveringParts}.
Now we have that
$C_{\xi'}(\varrho_L)-C_\xi(\varrho_L) = C_{\{X_{*1}',X_{*2}'\},X_*}(\varrho_{X_*})$, see \eqref{eq:deltaC}.
Since $\{X_{*1}',X_{*2}'\}=\{Y_{*1}',Y_{*2}'\}|_{Y_*}$, 
the proof is completed by \eqref{eq:restrictboundI}.
The meaning of \eqref{eq:valley} is exactly what we need:
if at a $\xi$ the change of the correlation $C_\xi$ is large
for all the possible steps then the change of the correlation
is also large if one leaves the up-set $\upset\{\xi\}$.
So if, 
during the successive refinement,
we follow a path in which the change of the correlation $C_\xi$ is small,
and we reach a $\xi$ of small enough $C_\xi$,
after which in every possible step this change becomes large, then we have reached $\beta$.

So in this way we have managed to give meaning to $C_{\xi'}-C_\xi$ being small or large,
by comparing the values of $C_\xi$ through a path from $\top$ to $\bot$.
But there is a question remained: 
how to do the successive refinement?
If, for example, we choose a wrong step \emph{in the beginning,} with $C_{\xi'}-C_{\xi=\top}=C_{\xi'}$ being large
(this is the case when $\beta\nfinereq\xi'$, we leave the up-set $\upset\{\beta\}$),
we do not notice this, and we miss the whole structure.
(Choosing a wrong step later can be recognised, 
since the difference $C_{\xi'}-C_\xi$ becomes large, compared to the differences in the previous steps.)
We can avoid this mistake if
in each step we choose the step in which $C_\xi$ changes the smallest.
However, always finding the step with minimal change in $C_\xi$ is still infeasible
($\abs{\{\xi'\in\Pi(L)| \text{$\xi'\covered\xi$}\}}=\sum_{X\in\xi} (2^{\abs{X}-1}-1)$ 
is still large in the beginning of the procedure).
Fortunately,
the bipartite correlation clustering $\gamma$, given in the previous section,
often gives us a good hint.
We can immediately have that
the parts $G\in\gamma$ should not be dissected:
the $2$-partitionability $C_{\text{$2$-part},G}$ in $G$ is strong \eqref{eq:edgeboundbondcorr2part},
which determines the change of $C_\xi$, that is,
\begin{equation}
T_\text{b}\leq C_{\xi'}-C_\xi,\qquad\text{if $\gamma\finer\xi$ and $\gamma\nfiner\xi'$,}
\end{equation}
see \eqref{eq:edgeboundbondcorr2part} and \eqref{eq:deltaC}.
This reduces the possibilities for the steps in the successive refining,
since it must be contained in $\upset\{\gamma\}$ until it reaches $\gamma$. 
However, in the presence of hidden correlations,
that is, strong multipartite correlations among the parts of $\gamma$,
we have that the change of $C_\xi$ is high even if no part of $\gamma$ is dissected
\eqref{eq:stepbound}.

%%%%%%%%%%%%%%%%%%%%%%%%%%%%%%%%%%%%%%%%%%%%%%%%%%%%%%%%%%%%%%%%%%%%%%%%%%%%%%%e
\subsection{Example for hidden correlations}

Here we construct an example family of states showing hidden correlations.
The smallest quantum system in which hidden correlations can occur is the system of three qubits.
A general three-qubit state $\varrho_{\{1,2,3\}}$ can be expressed in the basis of the Pauli matrices 
\begin{equation}
\sigma_0=\begin{bmatrix}1&0\\0&1\end{bmatrix},\;
\sigma_1=\begin{bmatrix}0&1\\1&0\end{bmatrix},\;
\sigma_2=\begin{bmatrix}0&-i\\i&0\end{bmatrix},\;
\sigma_3=\begin{bmatrix}1&0\\0&-1\end{bmatrix},
\end{equation}
with the coefficients $\varrho_{\{1,2,3\}}^{a,b,c} \in \field{R}$ as
\begin{equation}
\varrho_{\{1,2,3\}} = \frac18\sum_{a,b,c=0}^3\varrho_{\{1,2,3\}}^{a,b,c}\sigma_a\otimes\sigma_b\otimes\sigma_c.
\end{equation}
The normalisation $\tr\varrho_{\{1,2,3\}}=1$ leads to $\varrho_{\{1,2,3\}}^{0,0,0}=1$,
on the other hand, for the sake of simplicity, let us use only the $\sigma_0$ and $\sigma_3$ components,
that is, $\varrho_{\{1,2,3\}}^{a,b,c}=0$ if any of $a,b,c$ is $1$ or $2$.
Let then 
$\varrho_{\{1,2,3\}}^{3,0,0}=x$,
$\varrho_{\{1,2,3\}}^{0,3,0}=y$,
$\varrho_{\{1,2,3\}}^{0,0,3}=z$,
$\varrho_{\{1,2,3\}}^{0,3,3}=yz$,
$\varrho_{\{1,2,3\}}^{3,0,3}=xz$,
$\varrho_{\{1,2,3\}}^{3,3,0}=xy$,
and
$\varrho_{\{1,2,3\}}^{3,3,3}=v$.
For certain ranges of the parameters $(x,y,z,v)\in\field{R}^4$,
the resulting matrix is positive, that is, represents a state.
Then, using the notation $\varrho_{\{i\}}=\tr_{\{j,k\}}\varrho_{\{i,j,k\}}$, $\varrho_{\{i,j\}}=\tr_{\{k\}}\varrho_{\{i,j,k\}}$
for all distinct $i,j,k\in\{1,2,3\}$,
one can easily check that
\begin{subequations}
\begin{equation}
\varrho_{\{i,j\}} = \varrho_{\{i\}}\otimes\varrho_{\{j\}},
\end{equation}
while
\begin{equation}
\varrho_{\{1,2,3\}} = \varrho_{\{i\}}\otimes\varrho_{\{j,k\}}\quad\Longleftrightarrow\quad v=xyz.
\end{equation}
\end{subequations}
So, if $v\neq xyz$, then $\varrho_{\{1,2,3\}}$ is correlated with respect to any nontrivial split,
although its bipartite subsystems are uncorrelated.
Using the correlation measures \eqref{eq:xiCorr} and \eqref{eq:kCorrs},
this leads to
\begin{subequations}
\begin{align}
C_{i|j}(\varrho_{\{i,j\}}) &= 0,\\
C_{i|j,k}(\varrho_{\{1,2,3\}}) &> 0,\\
C_\text{$2$-part} (\varrho_{\{1,2,3\}}) &= 
C_\text{$2$-prod} (\varrho_{\{1,2,3\}}) > 0,\\
C_{1|2|3}(\varrho_{\{1,2,3\}}) =\qquad& \notag \\
=C_\text{$3$-part} (\varrho_{\{1,2,3\}}) &=
C_\text{$1$-prod} (\varrho_{\{1,2,3\}}) >0.
\end{align}
\end{subequations}

Important to note that $\varrho_{123}$ is a diagonal matrix,
the diagonal elements of which can be considered as the entries of a classical three-bit state.
That is, the phenomenon of hidden multipartite correlations is not a quantum feature,
it exists also in states of classical systems.

\clearpage
\onecolumngrid
%%%%%%%%%%%%%%%%%%%%%%%%%%%%%%%%%%%%%%%%%%%%%%%%%%%%%%%%%%%%%%%%%%%%%%%%%%%%%%%%
\section{Results employing the minimal basis}
Here we present the results of the same calculations as in the main text,
but now using STO-3G basis set.

\begin{figure}[b]
\centering
\includegraphics{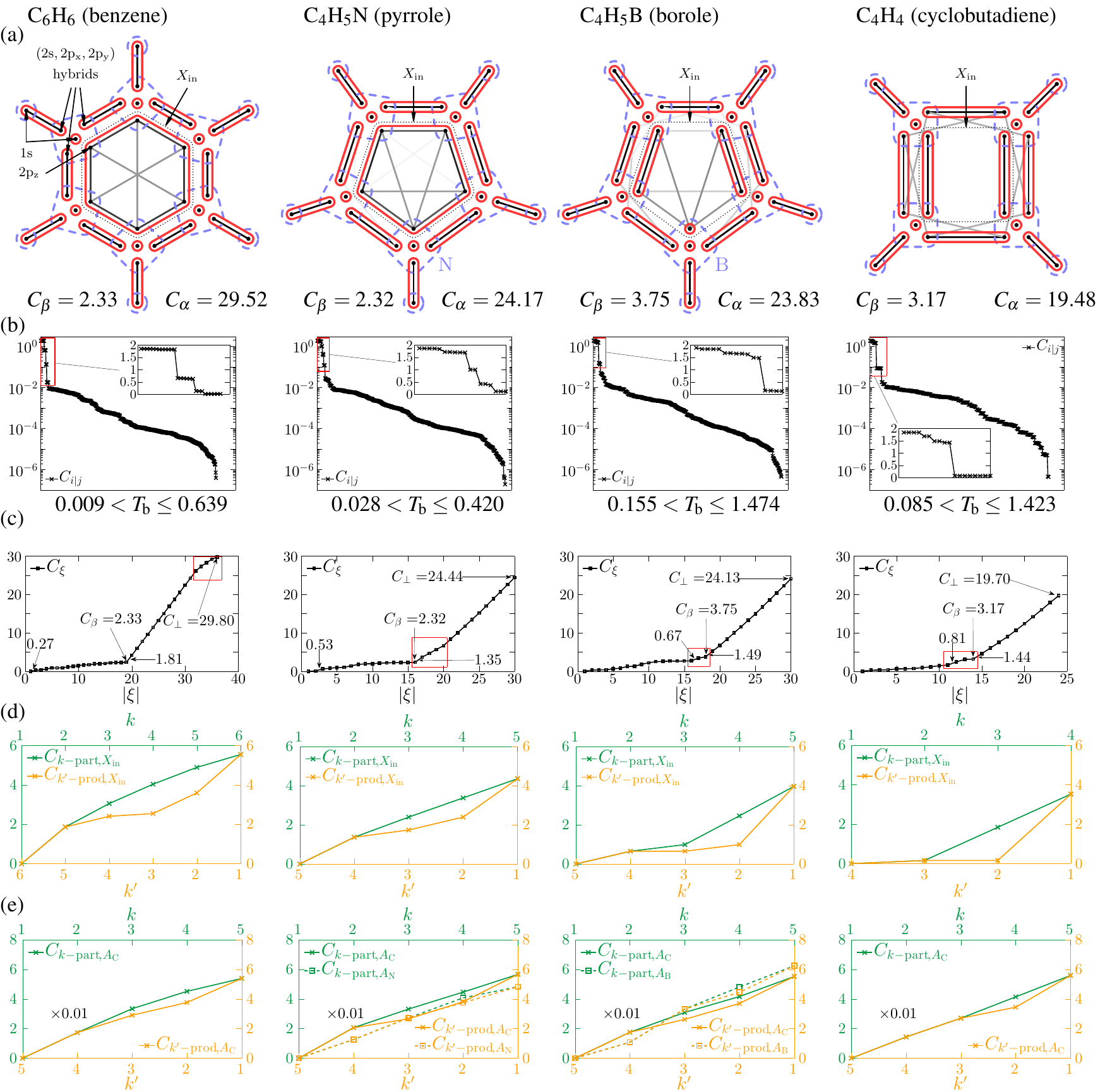}
\caption{Partitioning and multipartite correlations for the benzene, pyrrole, borole and cyclobutadiene molecules.
The same types of data are shown as in Fig.~1 in the main text.}
\end{figure}

\begin{figure}[b]
\centering
\includegraphics{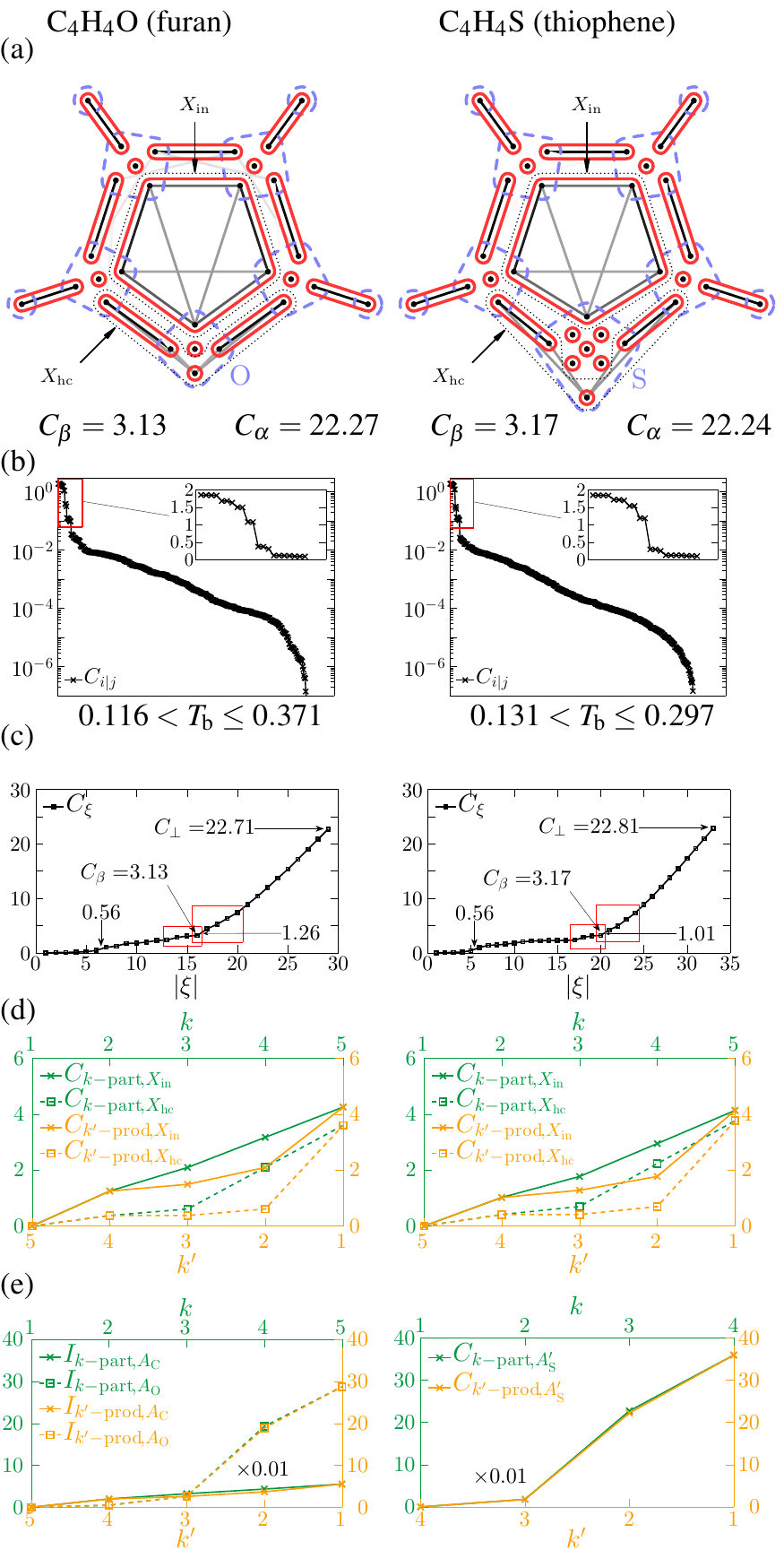}
\caption{Partitioning and multipartite correlations for the furan and thiophene molecules.
The same types of data are shown as in Fig.~2 in the main text.}
\end{figure}

\begin{figure}[b]
\centering
\includegraphics{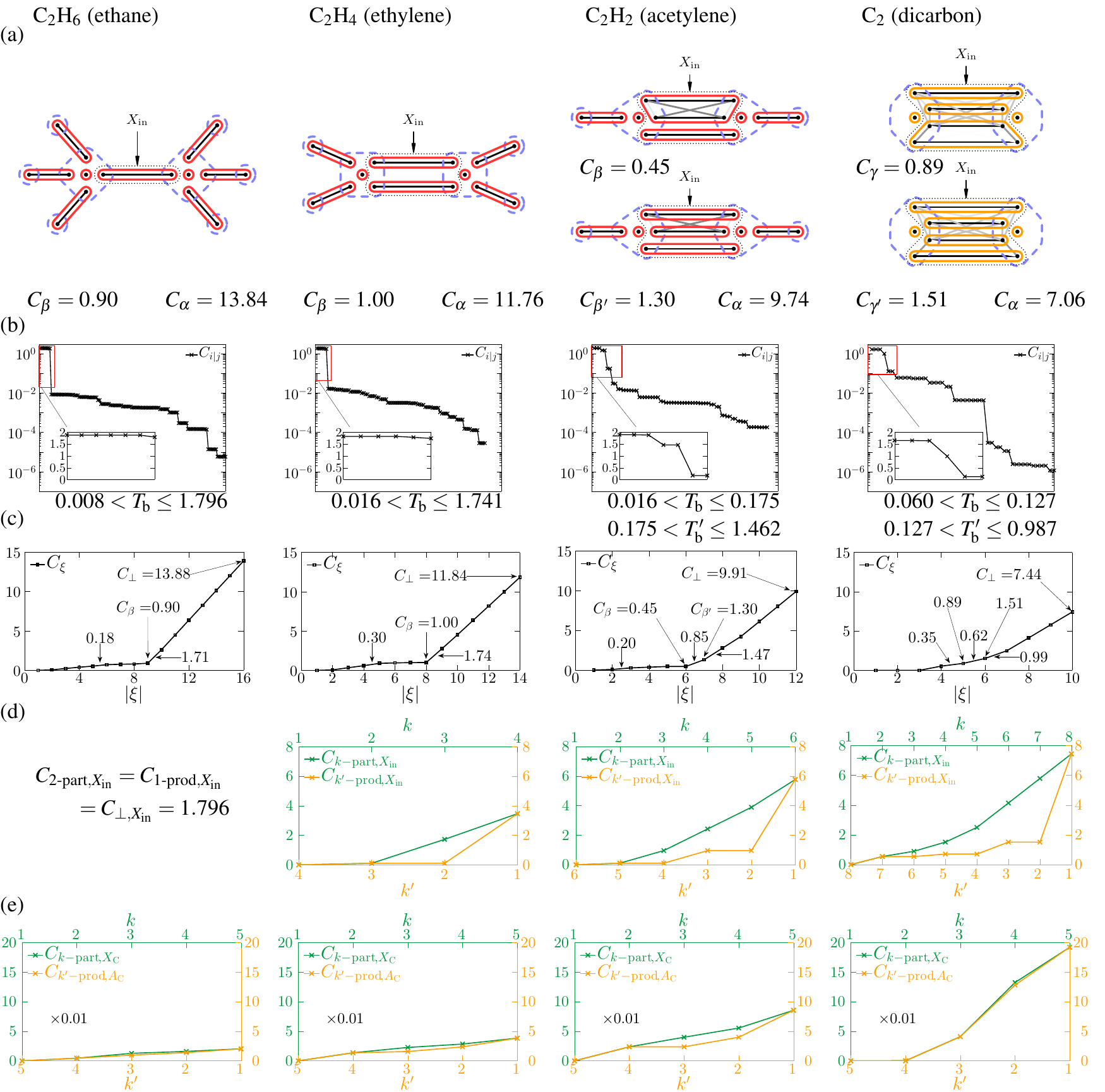}
\caption{Partitioning and multipartite correlations for the $\mathrm{C}_2\mathrm{H}_{2x}$ molecules.
The same types of data are shown as in Fig.~3 in the main text.}
\end{figure}

\clearpage
\twocolumngrid

\end{document}